\documentclass[useAMS,usenatbib]{mn2e}
\usepackage[utf8]{inputenc}
\usepackage{graphicx}
\usepackage{times}

\title[COORBITAL SATELLITES OF SATURN: CONGENITAL FORMATION]{COORBITAL SATELLITES OF SATURN: CONGENITAL FORMATION}
\author[ Izidoro, Winter \& Tsuchida]{A. Izidoro$^{1}$\thanks{E-mail: izidoro.costa@gmail.com (AI); ocwinter@pq.cnpq.br (OCW); tsuchida@ibilce.unesp.br (MT)}, O. C.
Winter$^{1}$\footnotemark[1] and M. Tsuchida$^{2}$\footnotemark[1]\\
$^{1}$ UNESP, Univ Estadual Paulista - Grupo de Dinâmica Orbital \& Planetologia, Guaratinguetá, CEP 12.516-410, Brazil\\
$^{2}$ UNESP, Univ Estadual Paulista - DCCE-IBILCE, São José do Rio Preto, CEP 15.054-000, Brazil\textit{}}
\begin{document}

\date{Accepted            Received; in original form  }


\maketitle

\label{firstpage}

\begin{abstract}

Saturn is the only known planet to have coorbital satellite systems.
In the present work we studied the process of mass accretion as a possible
mechanism for  coorbital satellites formation.
The system considered is composed of Saturn, a proto-satellite and a cloud 
of planetesimals distributed in the coorbital region around a triangular 
Lagrangian point. The adopted relative mass for the proto-satellite was $10^{-6}$ of Saturn's mass
and for each planetesimal of the cloud three cases of relative mass were considered, $10^{-14}$, $10^{-13}$ 
and $10^{-12}$ masses of Saturn. In the simulations each cloud of planetesimal was composed of $10^3$, $5\times 10^3$ or $10^4$ planetesimals. The results of the simulations show the formation of coorbital
satellites with relative masses of the same order of those found in the saturnian
system ($10^{-13}$ - $10^{-9}$). Most of them present horseshoe type orbits, but a
significant part is in tadpole orbit around $L_4$ or $L_5$. Therefore, the results indicate that this
is a plausible mechanism for the formation of coorbital satellites.

\end{abstract}


\section{Introduction}

Coorbital systems are those in which at least two bodies share the same mean orbit. Coorbital objects that librate around the Lagrangian points $L_4$ or $L_5$ are said 
to be in tadpole orbits, while those that librate around $L_4$, $L_3$ and $L_5$ are said to be in horseshoe type orbits. Although Lagrange have described the motion of bodies around these equilibrium 
points in 1788, when he published his {\it Analytical Mechanics}, only in 1906 a body showing this kind of motion was discovered by Max Wolf in Heidelberg.
He found an asteroid librating around the point $L_4$ of Jupiter in the system Sun-Jupiter. This asteroid is (588) Achiles and at the moment there are more than 3200 know asteroids
coorbiting with Jupiter (http:\\www.cfa.harvard.edu/iau/lists/JupiterTrojans.html). Apart from Jupiter there are other planets with coorbital bodies. In 1991 the asteroid
5261 Eureka was discovered librating around the $L_5$ of Mars (Innanen, 1991). Later three more asteroids were discovered: 1998 VF$_{31}$ (Tabachnik \& Evans, 1999), 1999 UJ${_7}$ (Connors et al., 2005) and 2007 NS$_2$ (http:\\www.cfa.harvard.edu/iau/lists/MarsTrojans.html). Neptune has six coorbital asteroids. All of them librating around its Lagrangian point $L_4$ (Zhou et al., 2008).

On the other hand, all the coorbital satellite systems known are around the planet Saturn (Table 1).The satellite Thetys has two coorbital satellites. Telesto around its $L_4$ point and Calypso around its $L_5$ point. They were discovered in 1980 by observations made from Earth (Veillet, 1981; Reitsema, 1981). The satellite Dione also has two coorbital satellites. Helene around its $L_4$ point and Polydeuces around its $L_5$ point. Helene was discovered in 1980 by observations made from Earth (Lecacheux et al., 1980; Reitsema et al., 1980) and Polydeuces was discovered through images from the Cassini spacecraft (Murray et al., 2005; Porco et al., 2005).
In 1981, images from Voyager not only confirmed the existence of the coorbital system Janus and Epimetheus but also determined their masses and orbital elements. In a suitable rotating coordinate system, both satellites perform horseshoe orbits and reach a distance of only 15000 km from each other (Yoder et. al, 1983).

\begin{table*}
\centering
 \begin{minipage}{100mm}
 \caption{ Coorbital Satellites of Saturn (\citet{b10}, Porco et al. (2005), Murray et al. (2005),Jacobson et al. (2004))}

\begin{tabular}{@{}lccccc@{}}
  \hline
 satellite & relative mass\footnote{Mass w.r.t. Saturn} & Orbit &  Density ($g/cm^3$) &Radius ($km$) & $\Delta\theta$\footnote{Libration Amplitude}  \\
  \hline\hline
Polydeuces      & 1$\sim5\times 10^{-13}$    & $L_5$ Dione   &  -   & 3.25 & 25.8     \\
\hline
Helene          & $4.48\times 10^{-11}$   & $L_4$ Dione   &  1.5   & $16\pm4$ & 14.8    \\
\hline
Telesto         &  $1.25\times 10^{-11}$     & $L_4$ Thetys    &1.0    & $12\pm3$ & 1.3     \\
\hline
Calypso         &  $6.32\times 10^{-12}$   & $L_5$ Thetys   & 1.0   & $9.5\pm1.5$ & 3.6    \\
\hline
Janus            & 3.38e-09  & horseshoe  & 0.65   & 99.3$\times$95.6$\times$75.6 &    \\
\hline
Epimetheus         &9.67e-10  & horseshoe  & 0.63   & 69$\times$55$\times$55 &    \\
\hline
\end{tabular}
\end{minipage}
\end{table*}

The satellites Thetys and Dione are of the same size and have mass of the order of $10^{-6}$ Saturn's mass (Table 2).
They also have another interesting common feature.
They are in mean motion resonance with a smaller interior satellite.
Mimas-Thetys are in 4:2 inclination type resonance while Enceladus-Dione in a 2:1 eccentricity type resonance.
However, Mimas and Enceladus do not have any coorbital satellite known until now.
A possible explanation for that was given by Mour\~ ao et al. (2006) and Christou et al. (2007).
Mourão et al. (2006) studied the stability of hypothetical satellites coorbital to the satellites Mimas and Enceladus. Their results showed that these coorbital satellites are in stable orbits. Then, they explored the stability of these hypothetical coorbital satellites putting Mimas and Enceladus at different values of semi-major axis, which they could have occupied in the past, along their orbital migration due to tidal their effects. They found that the hypothetical coorbital satellites would always be in stable orbits. The only exceptions occurred when Mimas and Enceladus were placed at semi-major axis where they were in mean motion first-order resonance with each other.
The hypothetical coorbital satellites left their tadpole orbits when Mimas and Enceladus were in 4:5, 5:6 or 6:7 mean motion resonances. Therefore, if there were coorbital satellites of Mimas or Enceladus, they may have lost them along their orbital migration, when passing through specific mean motion resonances.
\begin{table}
\centering
\begin{minipage}{70mm}
\caption{Satellites of Saturn which have coorbital satellites in tadpole orbits (\citet{b10})}

\begin{tabular}{@{}lccc@{}}
  \hline
        & \multicolumn{2}{c}{Satellites}\\
   satellite & relative mass\footnote{Mass w.r.t. Saturn} &  density ($g/cm^3$) & radius ($km$) \\
  \hline\hline
Dione          &$ 1.92\times 10^{-06}$    & 1.469   & 562.5     \\
\hline
Thetys          &$ 1.08\times 10^{-06}$   & 0.956   & 536.3     \\
\hline
\end{tabular}
\end{minipage}
\end{table}

Dermott \& Murray (1981a,b) presented a theory on the tadpole and horseshoe type orbits for the case
of circular and elliptic restricted three-body problem.
They also described the coorbital motion of Janus and Epimetheus through a numerical study combined with
perturbation theory. The study was made for a third body of negligible mass. Then, they generalized 
some of these results for the case when the third body disturbs the other two bodies of the system.
One of the main results found was a relation between the coorbital trajectory and its associated zero velocity curve.

There are several studies on the origin of the trojan asteroids, but not too many on the origin of the coorbital satellites. The origin of these satellites and Trojans asteroids are believed to be associated to one of the following mechanisms (Yoder et al.,1983; Smith et al.,1981): (i) capture by gas drag, (ii) congenital formation or (iii) rupture of a parent body.

Chanut et al. (2008) studied the mechanism of capture by drag.
They considered several values for the relative mass of the secondary body, from $10^{-7}$ to $10^{-3}$.
They explored the orbital evolution of planetesimals under orbital decay toward the secondary body orbit.
There are three possible outcomes along such evolution: collision with the secondary body, 
capture in the coorbital region or crossing the coorbital region toward the central body.
Once a planetesimal is captured in a coorbital trajectory, its amplitude of oscillation around one of the triangular Lagrangian points shrinks until it is fixed over the Lagrangian point. They also showed that the location of the Lagrangian points change according to the density of the gas.

With the numerous discoveries of the extrasolar planets Beaugé et al. (2007) analyzed the formation of hypothetical Earth-type coorbital planets in extrasolar systems. The central body was taken as equal to one solar mass and the giant planet with Jupiter's mass. They considered two scenarios of formation, one gas-free scenario and another rich-gas scenario. In both of them there were the formation of a single coorbital of the terrestrial type, but due to gravitational instability with other bodies, the accretion process was not efficient.

In this paper we study the mechanism of congenital formation of coorbital satellites and in some points this mechanism is similiar to adopted in Beaugé et al. (2007). According to the modern theories of planetary formation, in the beginning of the solar system formation there was a cloud of gas and dust that settled in a disk.  
Due to several factors the gas dissipated along the time.
In the first stages the particles were of the size of sub-micrometers up to the order of centimeters. 
This disk of gas and dust gradually generated larger bodies through physical collisions leading to agglomeration.
Successive encounters among them resulted in a large number of meter sized aggregates. From these objects originated planetesimals of the order of kilometers. 
Further encounters between planetesimals led to the formation of the planet's progenitors of terrestrial size or even ten times this size (Armitage, 2007). 
The formation of planetary satellite systems are thought to have followed a very similar way (Safronov, 1969; Wetherill 1980, 1990; Hayashi et al. 1985; Greenberg, 1989). 

In the present work we consider an intermediate stage of Saturn's satellite system formation.
In this stage all the major satellites are almost formed (here called proto-satellites), there is almost no gas left, but there is still plenty of small planetesimals in the disk.
Our model of study is characterized by a central body (Saturn), a secondary body (proto-satellite) and a cloud of plantesimals. Since it is assumed that the proto-satellite has already a significative mass, which is several orders of magnitude larger than the planetesimals, it defines (together with the planet) the structure of the phase space with the five Lagrangian points. Then, we consider a cloud of planetesimals in the coorbital region around one of the triangular Lagrangian points, under the gravitational influence of Saturn and the proto-satellite. The collision between planetesimals are considered constructive and through this accretion process larger bodies are built in the coorbital region.

This paper has the following structure. 
In the next section we present the methodology and assumptions adopted in our numerical simulations.
The results of the simulations are presented in section 3.
Finally, in the last section we present some final comments and our conclusions.

\section{Methodology}
The dynamical system considered is composed by Saturn (the central body), a proto-satellite and a cloud of planetesimals.
The mass of the proto-satellite was chosen to be $10^{-6}$ of Saturn's mass, which is close to the masses of Dione and Thetys (Table 2). It is placed in a circular orbit with the semi-major axis equal to that of Thetys. The mass of the proto-satelite might vary along the simulation due to some collisions with planetesimals, but its small growth do not affect significantly the dynamic of the planetesimals around the Lagrangian points (Fleming and Hamilton, 2000).

The planetesimal clouds were always initially randomly distributed in a sector around $L_4$ or $L_5$.
The sector is delimited by an arc of $80\deg$, centered on the Lagrangian point, and the extreme orbital radii of the largest horseshoe orbit (Figure 1). The half width of the largest horseshoe orbit is given by Dermott \& Murray (1981a) as
\begin{equation}
 \Delta {\bf r}_{\rm{horse}}=\frac{1}{2}\mu_2^{1/3}a_{2}
\end{equation}
where $\mu_2$ and $a_2$ are the relative mass and the semi-major axis of the proto-satellite, respectively. Studying the overlap of first order mean motion resonances, Wisdom (1980) found an expression that gives the width of a chaotic region from the planar circular orbit of a secondary body. Therefore, it is important to have in mind that just outside the coorbital region (interior and exterior) there is a chaotic region. Particles in this region usually tend to have close encounters with the proto-satellite and get large eccentricities, leaving the neighborhood.
\begin{figure}
\hspace{-0.5cm}
\includegraphics[scale=.5]{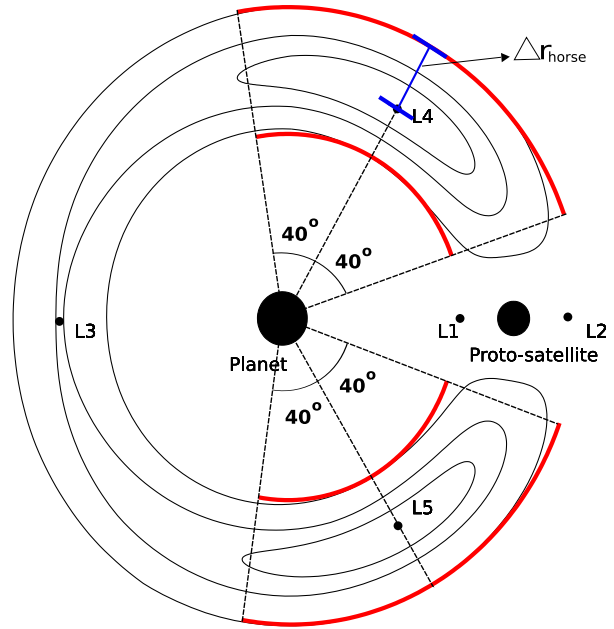}
\vspace{0.5cm}
\caption{Schematic location of the sectors around $L_4$ and $L_5$ were the planetesimals are initially distributed.
Each sector is delimited by an arc of $80\deg$, centered on the Lagrangian point, and the extreme orbital radii of the largest horseshoe orbit. The half width of the largest horseshoe orbit, $\Delta {\bf r}_{\rm horse}$, is given by Equation (1).
}
\end{figure}
Since we are interested in the formation of the small counterparts of coorbital satellite systems, in general, we considered clouds of planetesimals with the total mass of the order of the smaller coorbital satellites, given in table 1. 
We also considered that all planetesimals have the same initial mass,  $m_{p_i}$.
We performed simulations for different clouds of planetesimals.
We adopted clouds of $10^3$, $5\times 10^3$ and $10^4$ planetesimals
each one for three different values of $m_{p_i}$, $10^{-12}$, $10^{-13}$ and $10^{-14}$ masses of Saturn.

The simulations were made through numerical integrations using the Burlish-Stoer integrator from the package Mercury (Chambers, 1999). The length of the integration, $\tau$ (in orbital periods of the proto-satellite), varied according to the value of $m_{p_i}$. For $m_{p_i}=10^{-12}$ it was used $\tau=10^5$,
for $m_{p_i}=10^{-13}$ it was used $\tau=10^6$ and for $m_{p_i}=10^{-14}$ it was used $\tau=10^7$.
That is necessary due to the gravitational attraction between the planetesimals, 
which is smaller as smaller are their masses, leading to a longer time for the system to evolve.

For each cloud of planetesimals with a specific  value of $m_{p_i}$ we performed at least two independent simulations. One placing it around $L_4$ and the other around $L_5$. We also tested the effect of Saturn's oblateness by running extra simulations with the inclusion of the $J_2$ term in the gravitational potential of the central body. The planetesimals in all simulations were initially placed in circular orbits. 

\section{Numerical Simulations}

Along the numerical simulations the planetesimals collide among them generating lager planetesimals. The collisions are always considered constructive. i.e. inelastic collisions. By the end of the integration time there are only a few (at most four) large planetesimals left. The final masses of the remaining planetesimals  are represented by $m_{p_f}$. 

In order to have an idea of the time evolution of these systems we present a  sample of snapshots for two representative simulations.
Figure 2 shows the dynamical evolution of a cloud of $10^4$ planetesimals with $m_{p_i}=10^{-13}$, initially distributed around $L_4$. In each frame are shown the values of the osculating semi-major axis and eccentricities of the remaining planetesimals at a given moment. The orbital radius of the proto-satellite is considered as unity. The color grade indicates the mass of the planetesimal. 

Since we are interested in the formation of small coorbital satellites, the best candidates should be the remaining planetesimals with low eccentricity and semi-major axis inside the coorbital region. So we included in the plots the dashed curves that indicate the coorbital region defined by the largest horseshoe width ($1\pm \Delta {\bf r}_{ \rm horse}$). In general, the planetesimals that leave the coorbital region present chaotic behavior and collide with the proto-satellite.

From the plots of Figure 2 we notice that the eccentricities of the planetesimals increase as they start to grow. However, they are limited to low values ($e<0.007$). After about 1068 years there are only three planetesimals left and it takes about a thousand years to reach a single planetesimal. That last one has a mass of $m_{p_f}=5.1170E-10$ 

In Figure 3 we present the dynamical evolution of a cloud of $10^3$ planetesimals with $m_{p_i}=10^{-14}$, initially distributed around $L_5$. The evolution is very much similar to that presented in Figure 2. However, at least two major differences should be pointed out. The eccentricities do not get much larger then 0.001 and the time scales are much longer then in the previous case. These two features are naturally due to the lower values of the initial mass of the planetesimals, $m_{p_i}$, and the total mass of the cloud of the planetesimals. Larger masses produce stronger gravitational interactions accelerating the evolution of the system and increasing the values of eccentricities. 

\begin{figure*}
 \begin{minipage}{170mm} 
\centering
\includegraphics[width=0.43\linewidth]{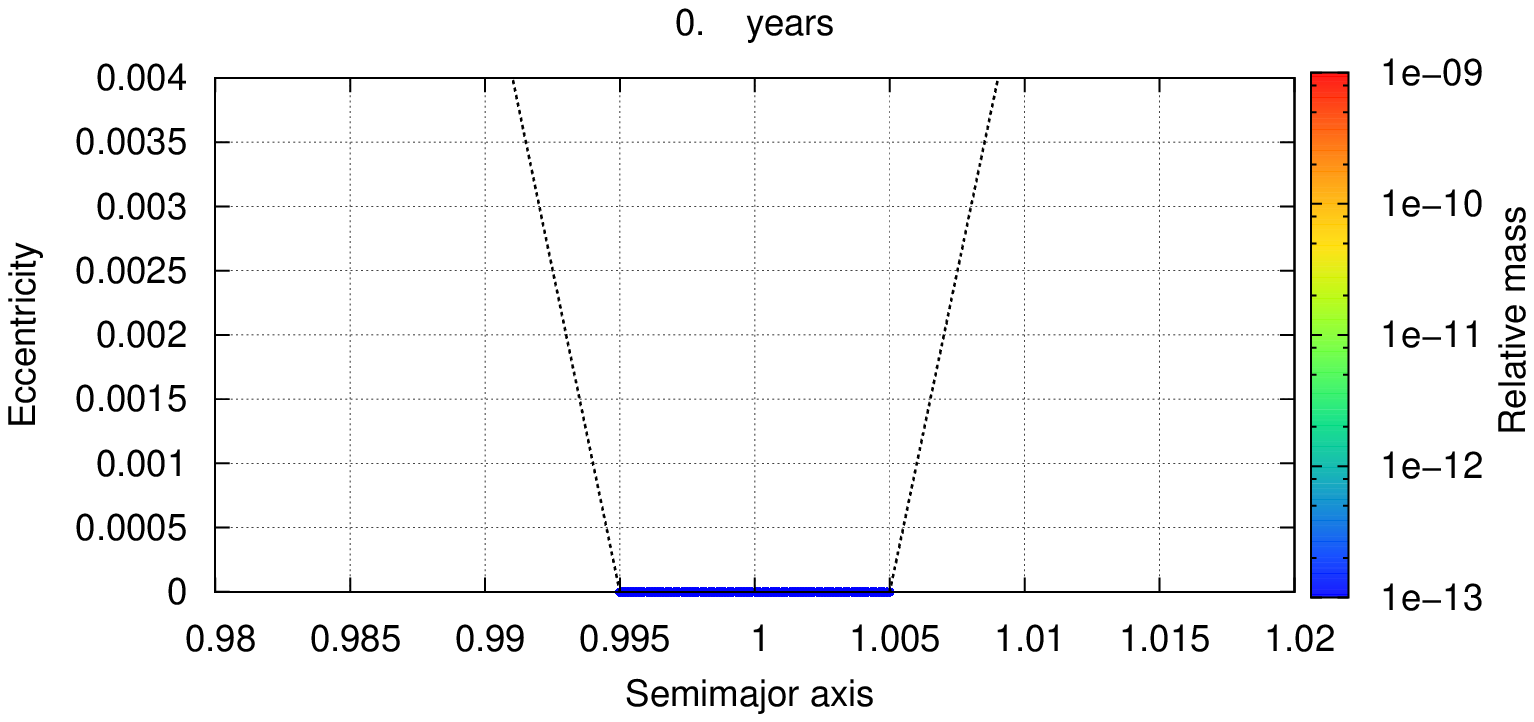}
\includegraphics[width=0.43\linewidth]{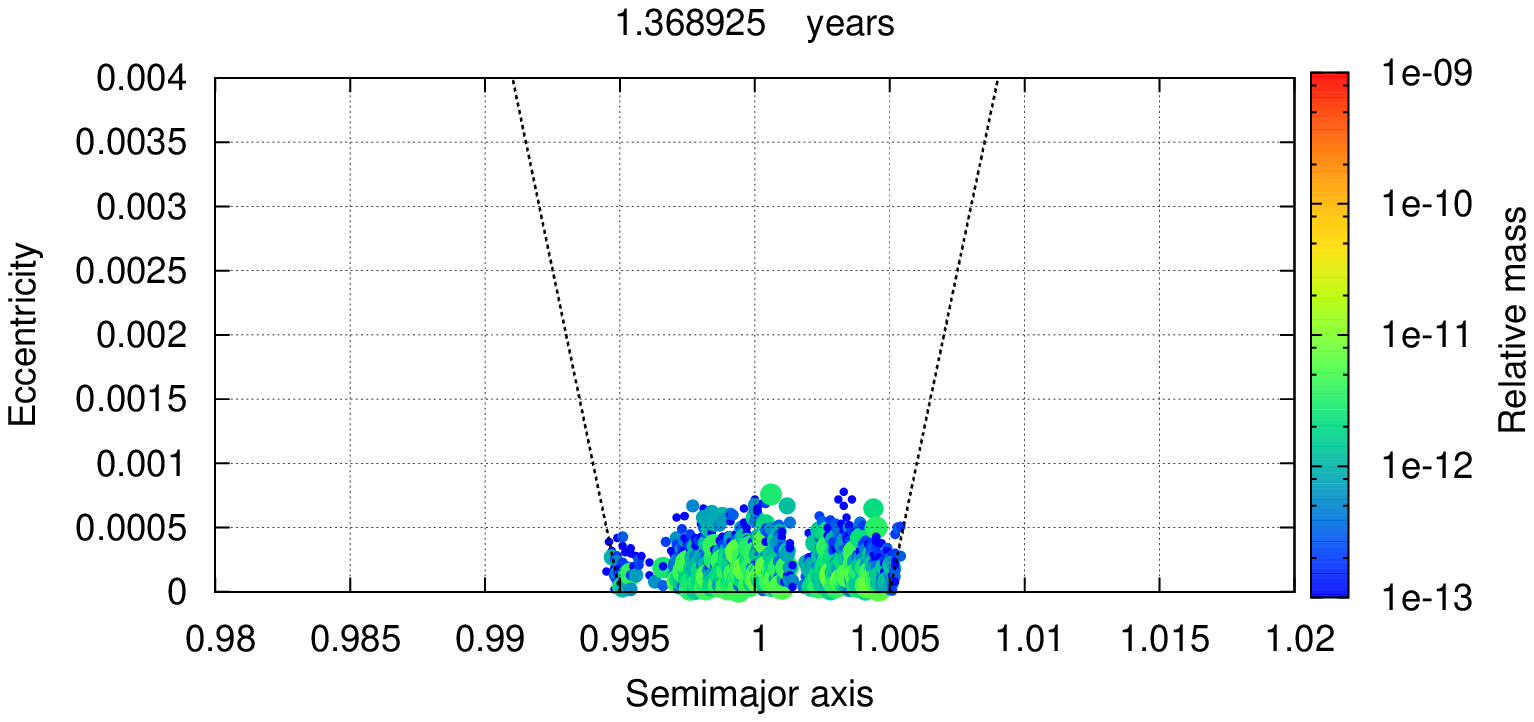}
\includegraphics[width=0.43\linewidth]{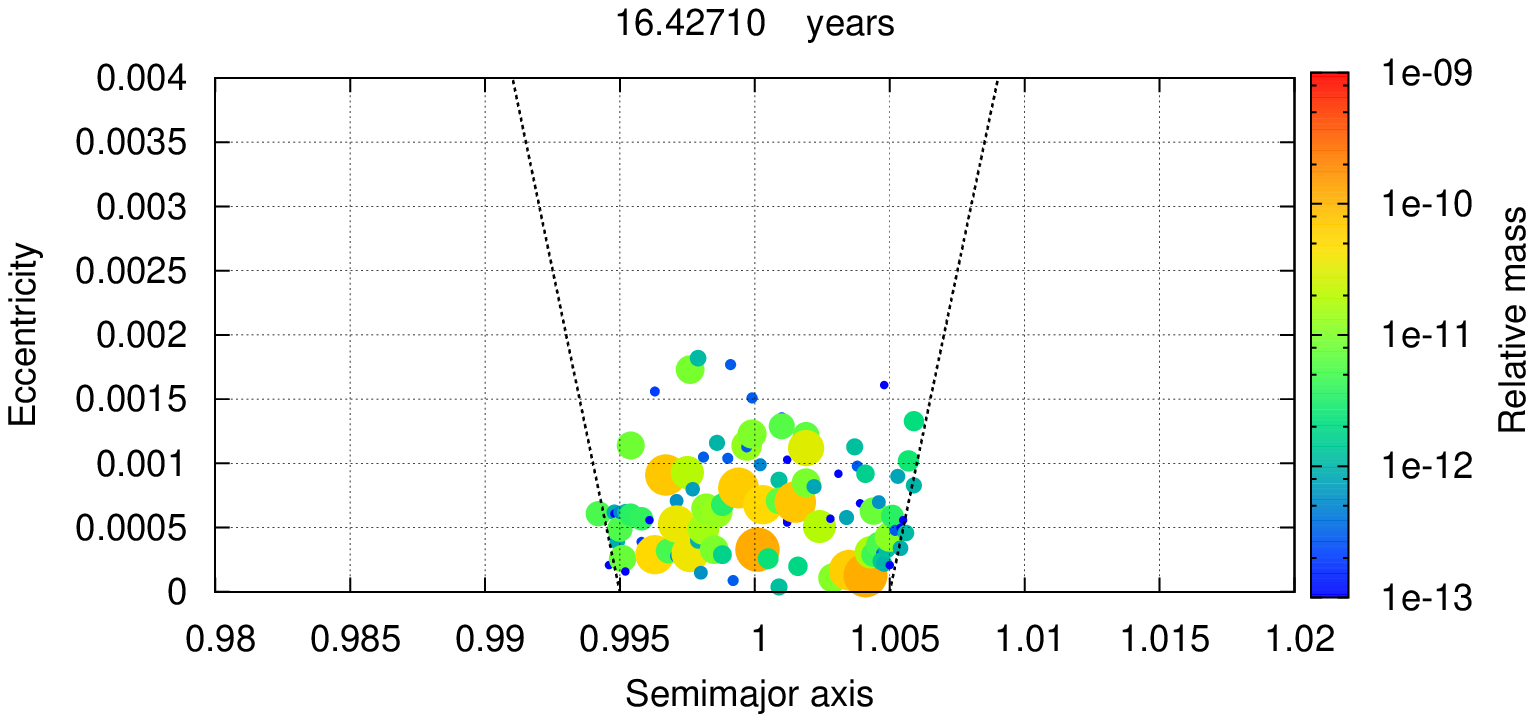}
\includegraphics[width=0.43\linewidth]{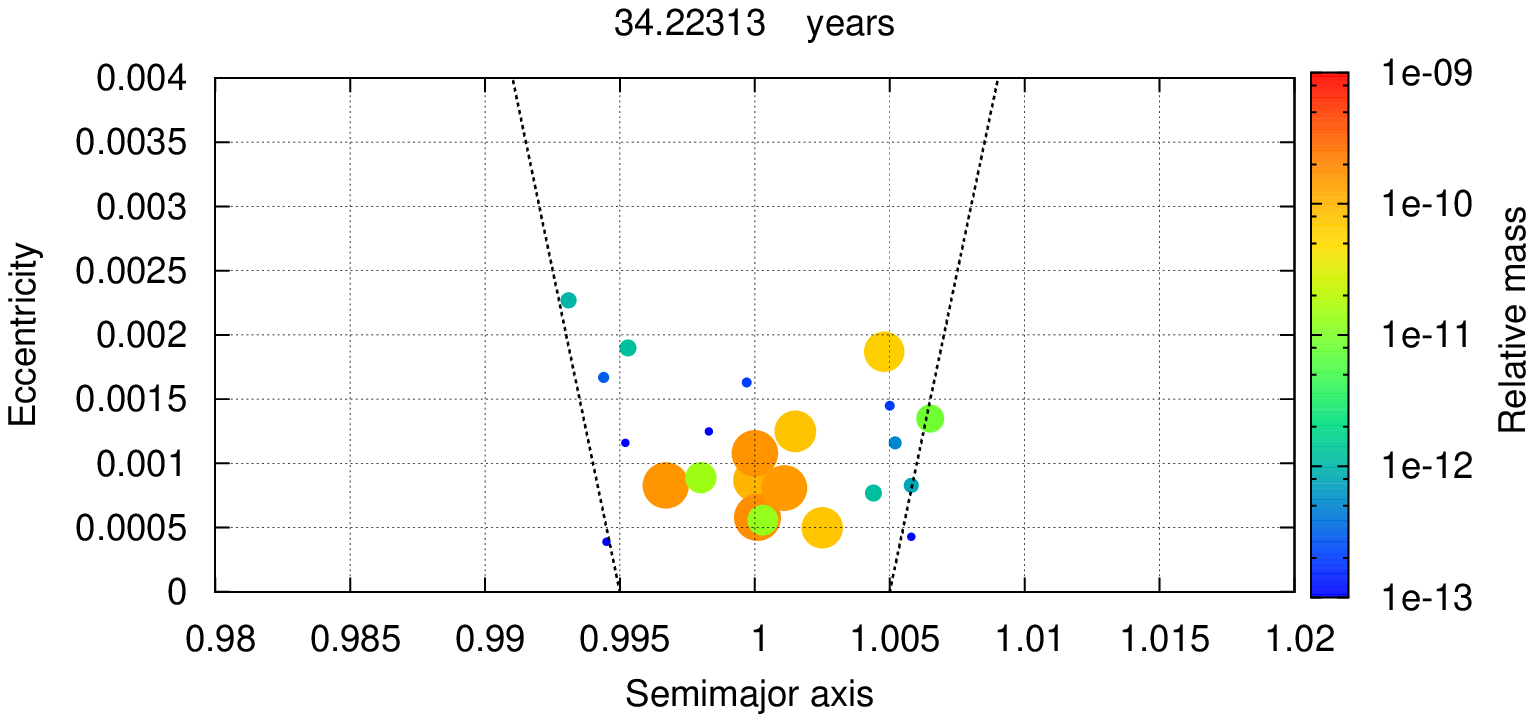}
\includegraphics[width=.43\linewidth]{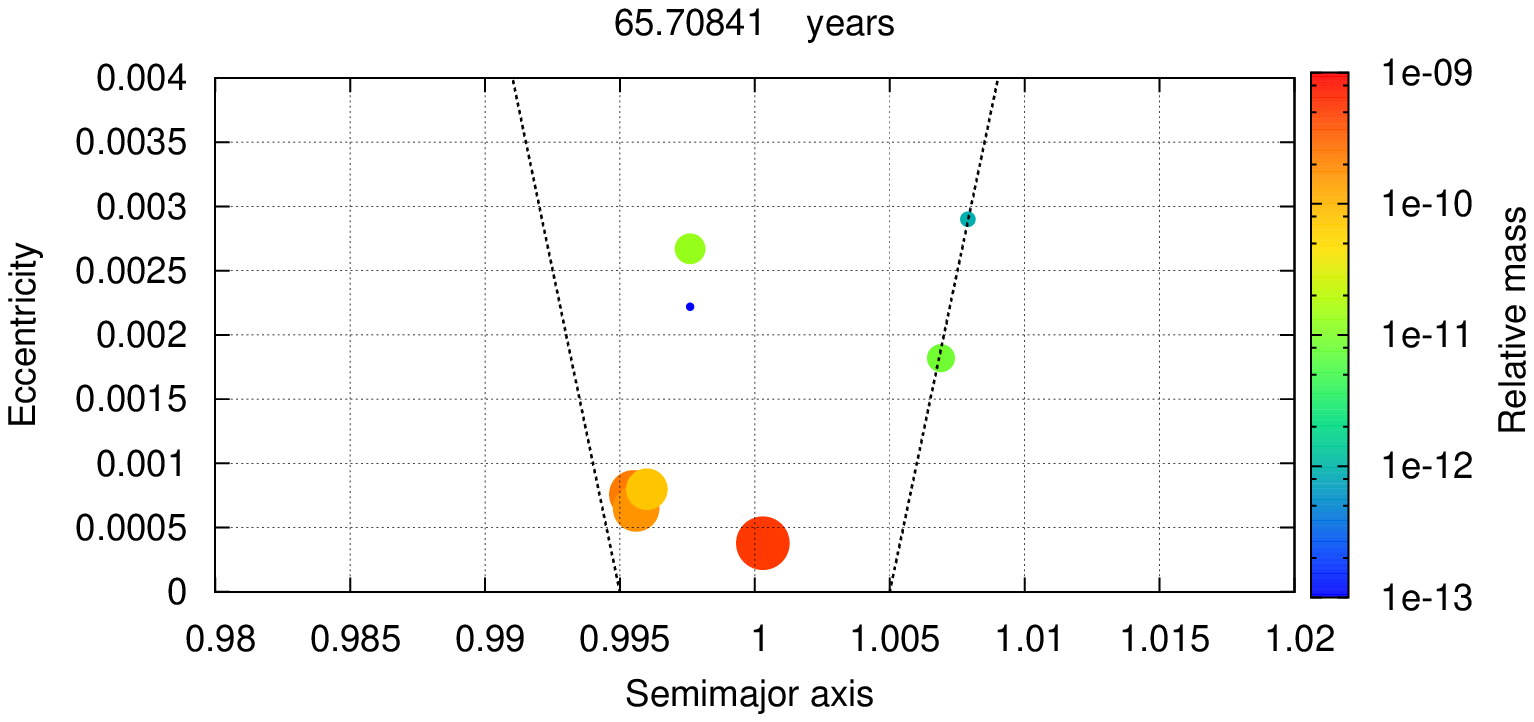}
\includegraphics[width=.43\linewidth]{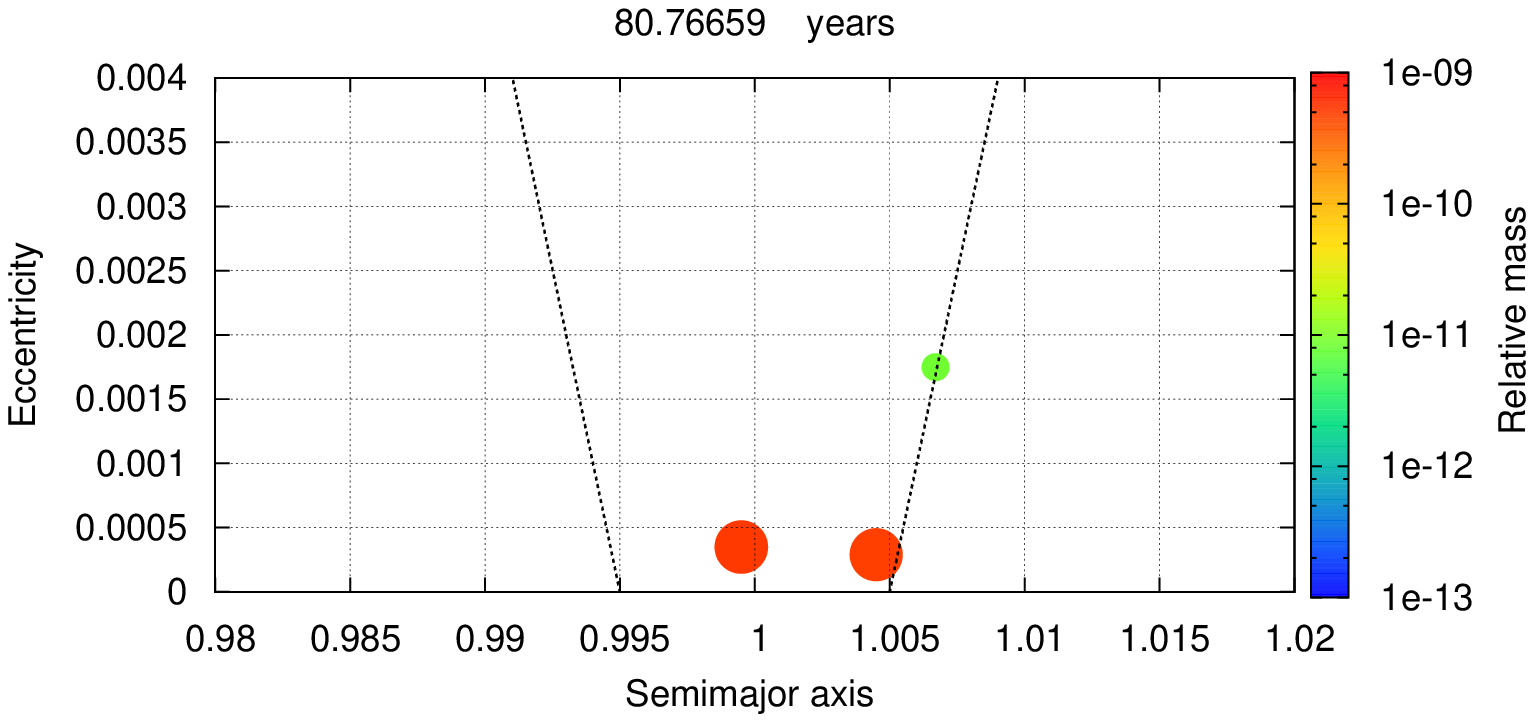}
\includegraphics[width=.43\linewidth]{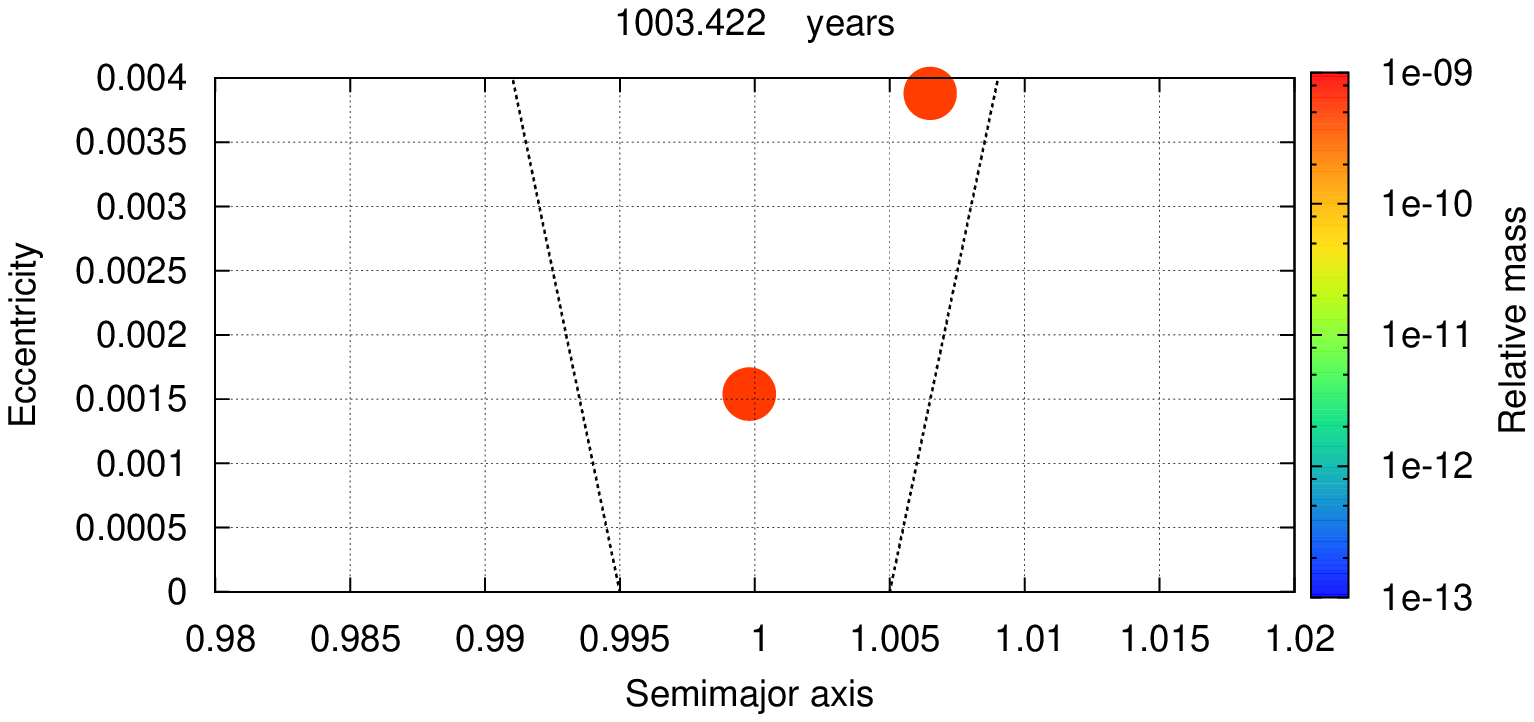}
\includegraphics[width=.43\linewidth]{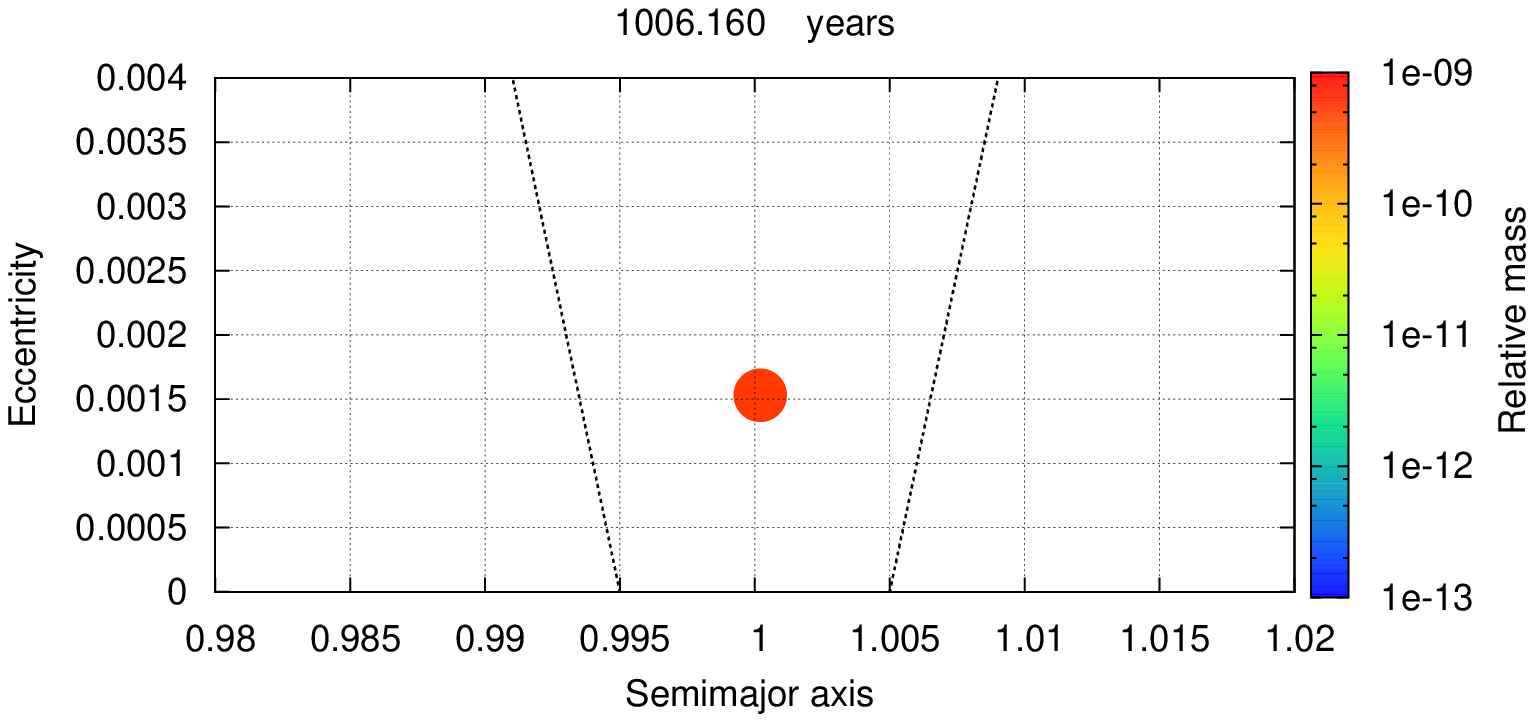}
\vspace{1cm}
\label{figura 1}
\caption{Dynamical evolution of a cloud of $10^4$ planetesimals with $m_{p_i}=10^{-13}$, initially distributed around $L_4$. In each frame are shown  the values of the osculating semi-major axis and eccentricities of the remaining planetesimals at a given moment. The color grade indicates the mass of the planetesimal. The dashed curves indicate the coorbital region defined by the largest horseshoe width ($1\pm \Delta {\bf r}_{\rm horse}$).
}
\end{minipage}
\end{figure*}

\begin{figure*}
 \begin{minipage}{170mm} 
\centering
\includegraphics[width=0.43\linewidth]{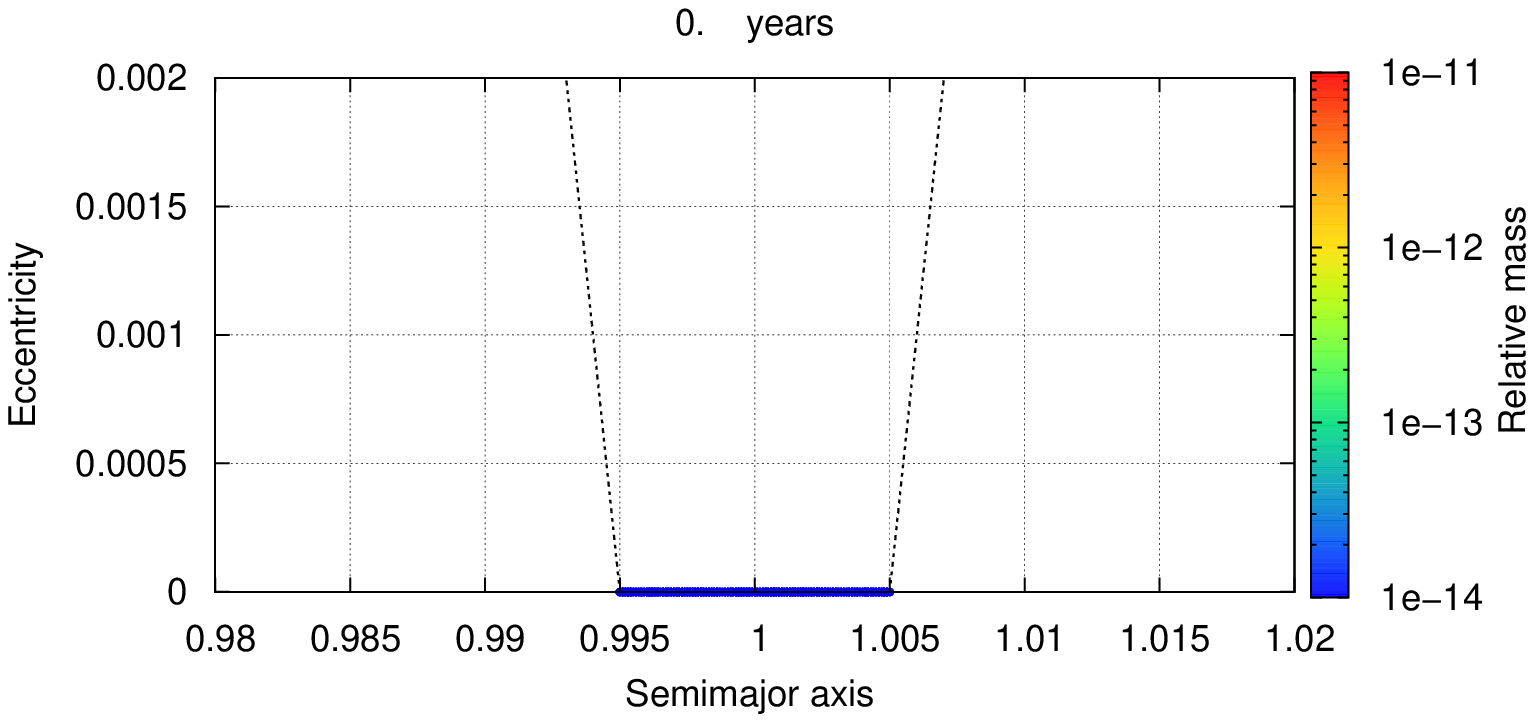}
\includegraphics[width=0.43\linewidth]{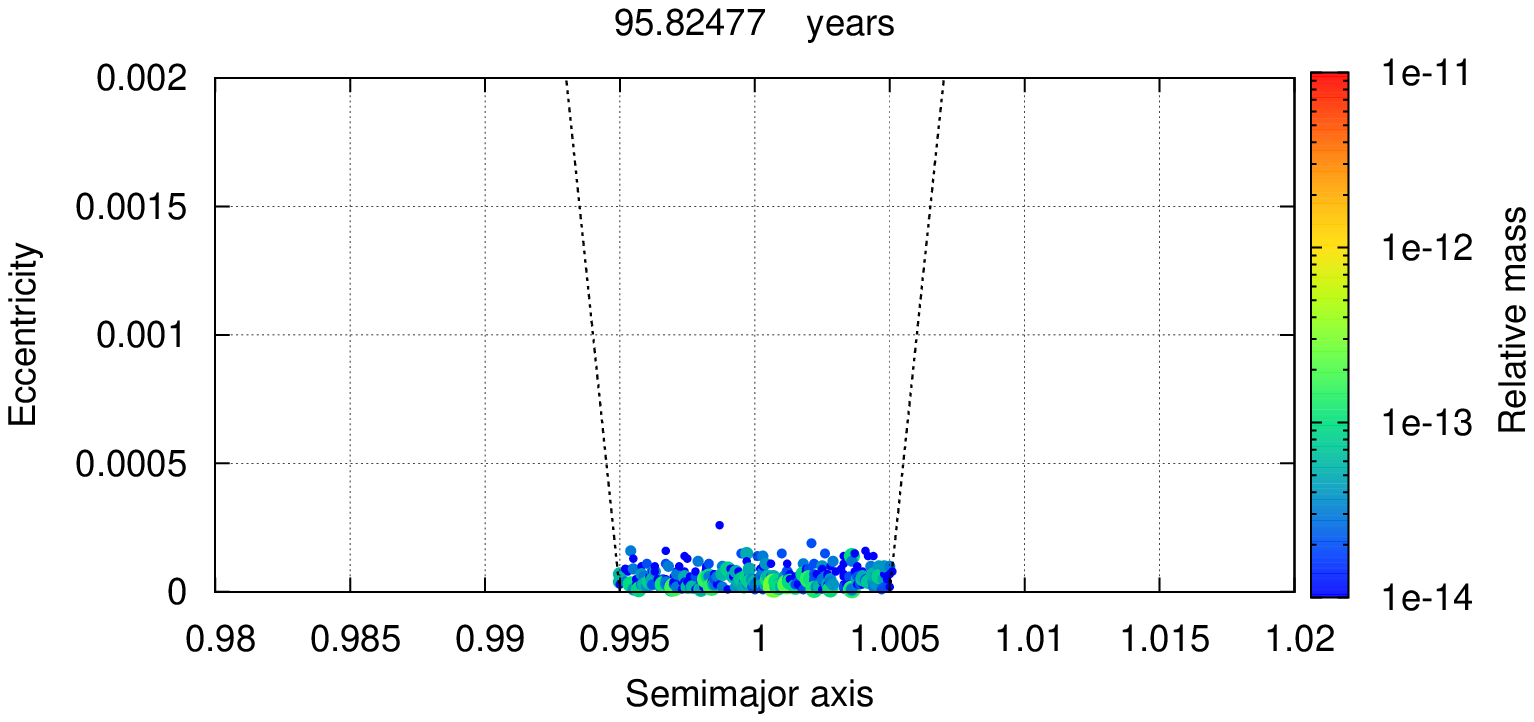}
\includegraphics[width=0.43\linewidth]{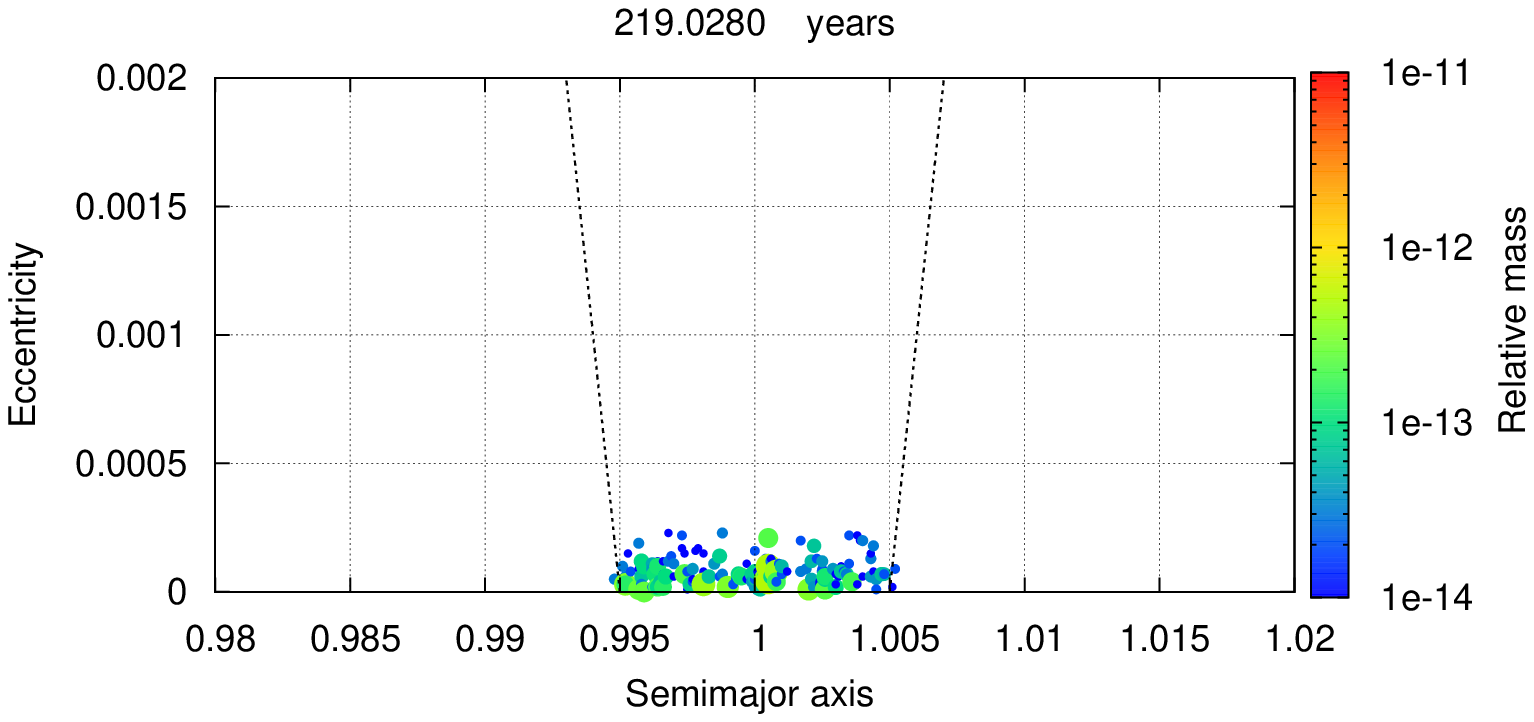}
\includegraphics[width=0.43\linewidth]{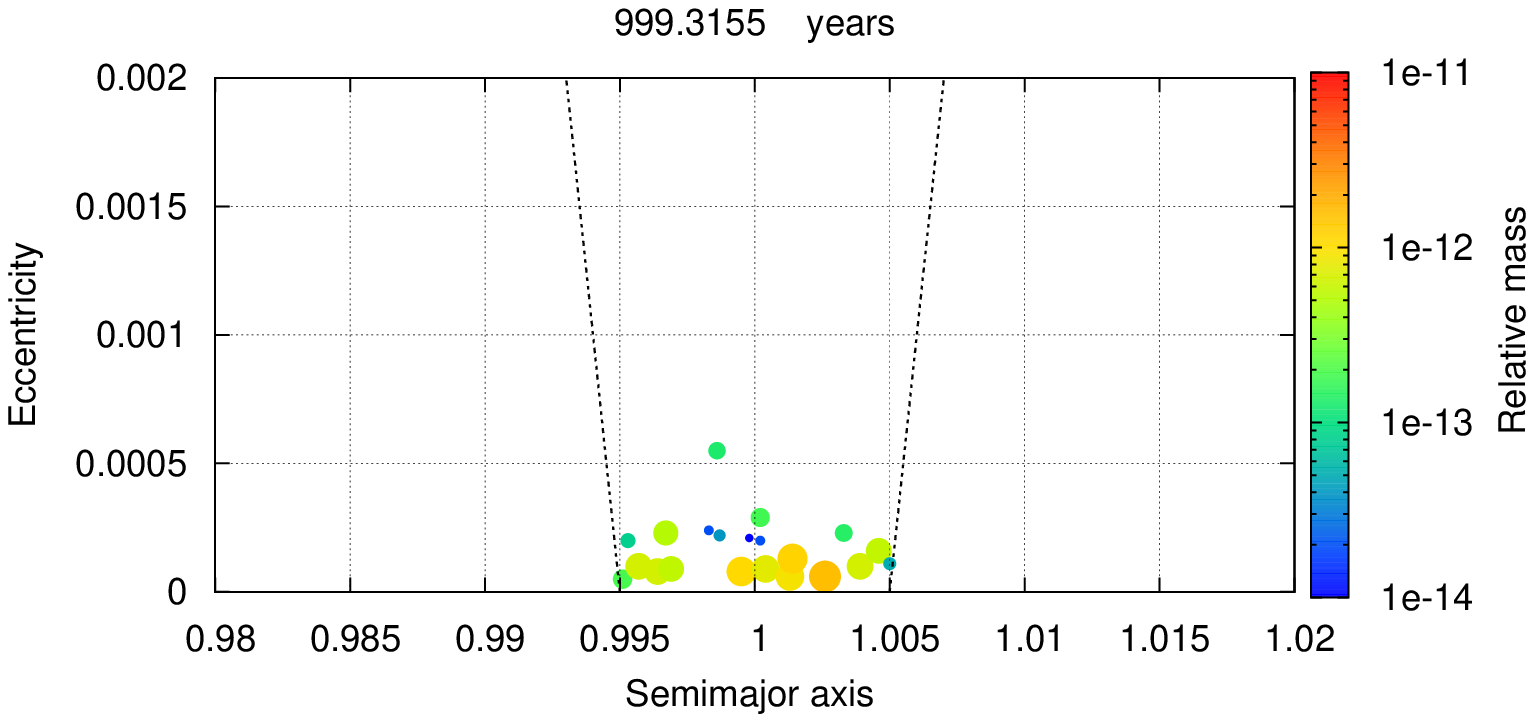}
\includegraphics[width=.43\linewidth]{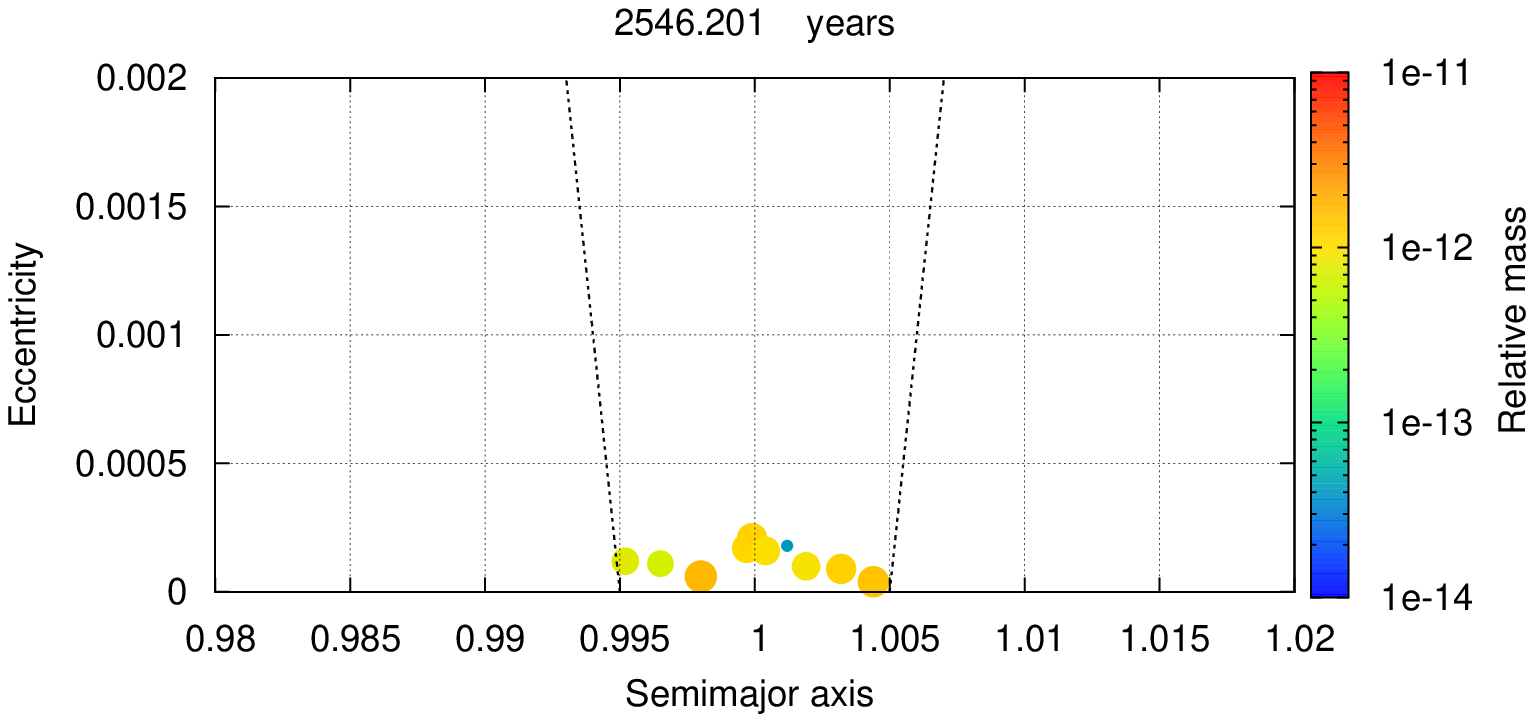}
\includegraphics[width=.43\linewidth]{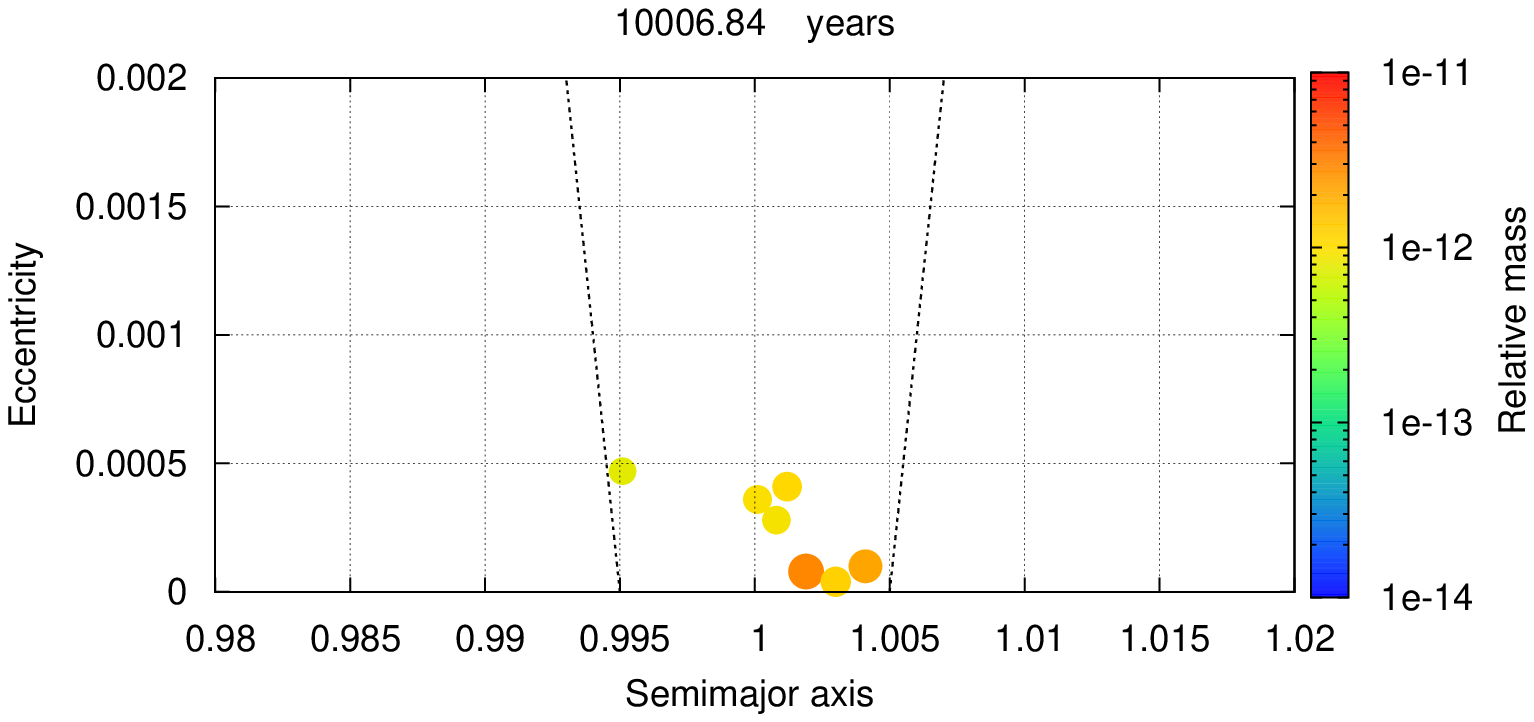}
\includegraphics[width=.43\linewidth]{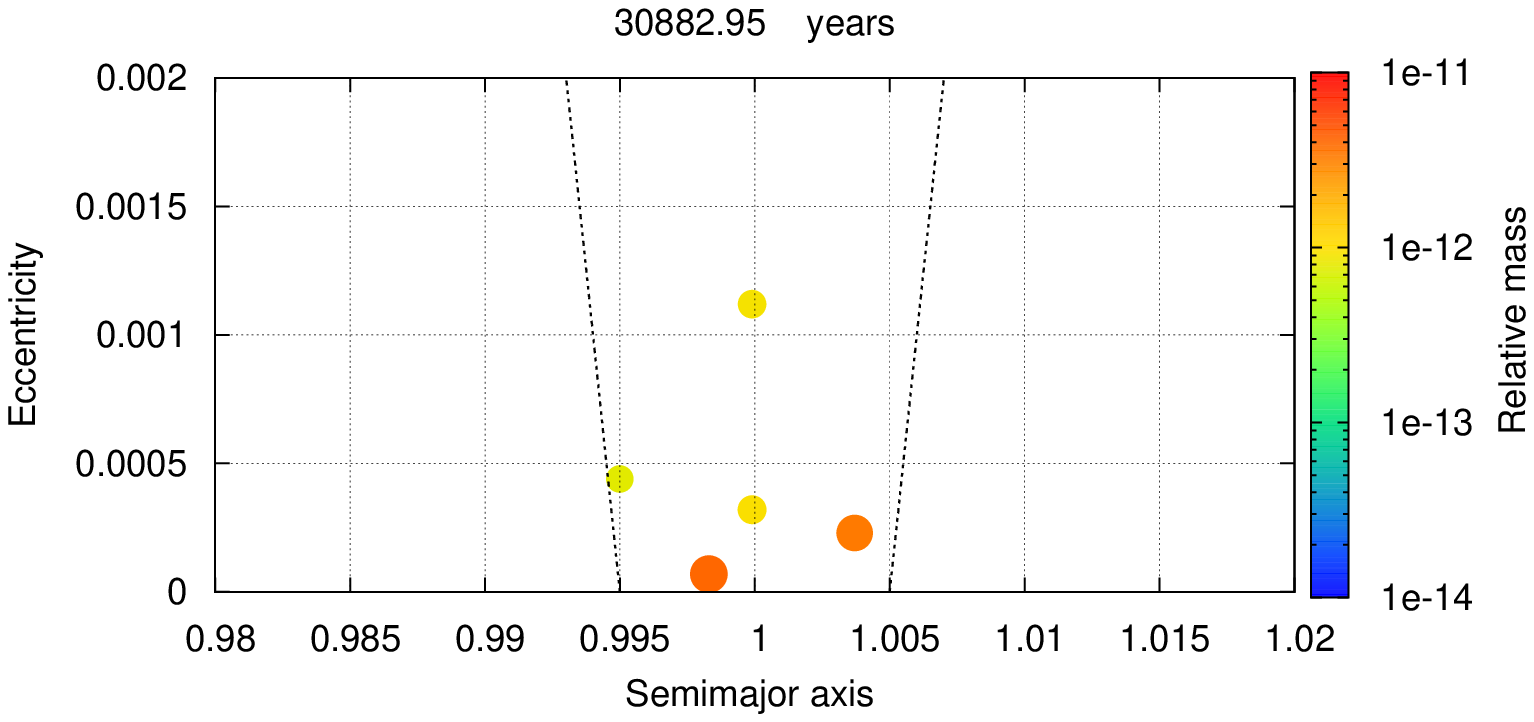}
\includegraphics[width=.43\linewidth]{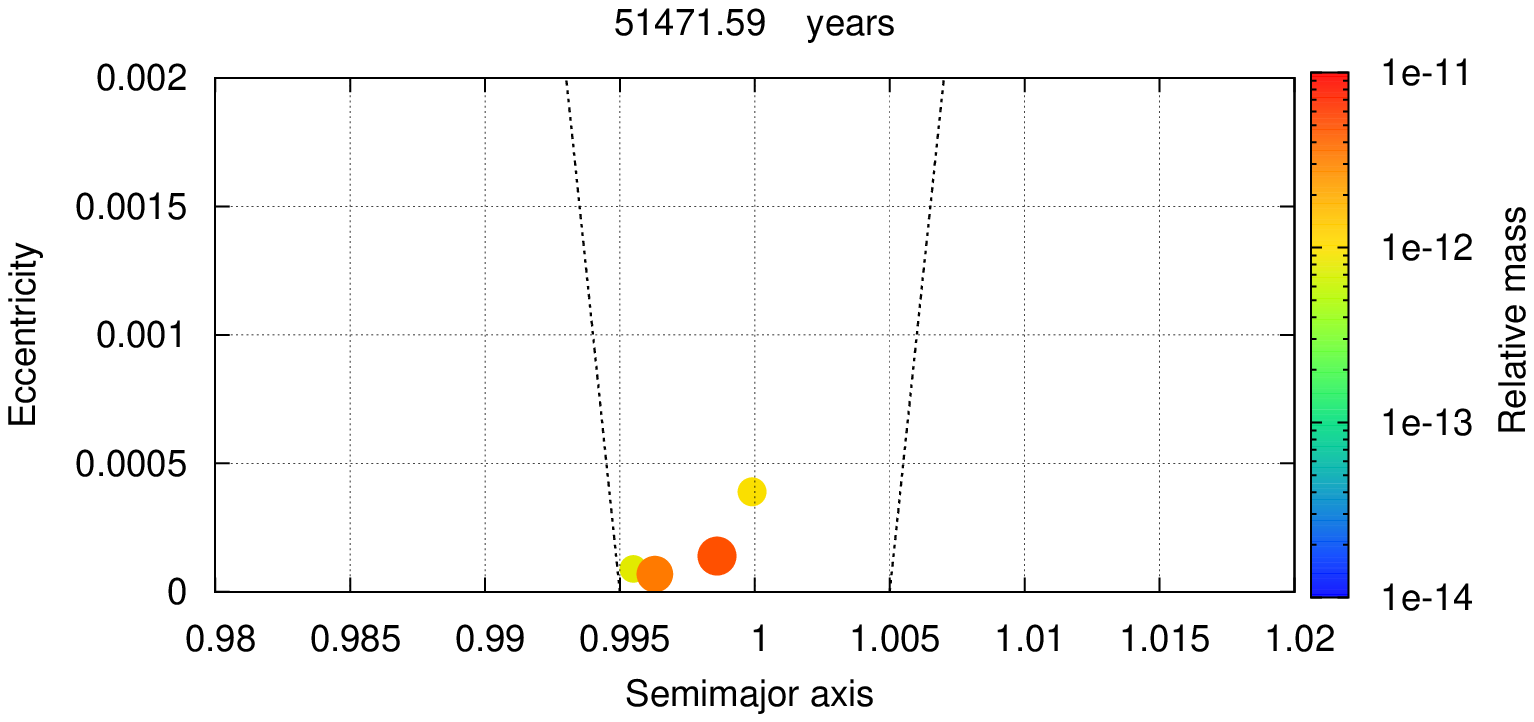}
\vspace{1cm}
\caption{Dynamical evolution of a cloud of $10^3$ planetesimals with $m_{p_i}=10^{-14}$, initially distributed around $L_5$. In each frame are shown  the values of the osculating semi-major axis and eccentricities of the remaining planetesimals at a given moment. The color grade indicates the mass of the planetesimal. The dashed curves indicate the coorbital region defined by the largest horseshoe width ($1\pm \Delta {\bf r}_{\rm horse}$).
}
\end{minipage}
\end{figure*}

The evolution of all the other numerical simulations we have performed are similar to these ones. However, the plots presented in Figures 2 and 3 do not guarantee that the remaining planetesimals are in coorbital orbits with the proto-satellite. Therefore, we analyzed the evolution of the angle $\theta$ between the remaining planetesimals and the proto-satellite, called libration angle. This angle is defined as the difference in mean longitudes, $\theta=\lambda_p-\lambda_2$, where $\lambda_p$ and $\lambda_2$ are the mean longitudes of the planetesimal and secondary body respectively.
A representative sample of such evolution is presented in Figures 4 to 7.
In each figure we present two plots. The top plot shows the evolution of $\theta$ and the bottom one shows the evolution the of planetesimal's mass. All plots cover the
whole integration time and they are shown in a logarithmic scale.
The temporal evolution of the libration angle confirmed that in all simulations the remaining planetesimals are in coorbital trajectories. The plots show a complete variety of evolutions. Planetesimals that started in tadpole orbits around $L_4$ or $L_5$ and ended in horseshoe orbits (Figures 5 and 6). Planetesimals that started in tadpole orbit around $L_4$ and ended in tadpole orbit around $L_5$ (Figure 4) or in tadpole orbit around $L_4$ (Figure 7). The evolutions are not always smooth. There are cases of tadpole orbits that "jump" from $L_4$ to $L_5$ and then again back to $L_4$ (Figure 7). And others that "jump" from $L_5$ to $L_4$ and then to a horseshoe orbit (Figure 6).

The changes in the amplitude of oscillation of the coorbital planetesimals along their evolution are associated to the collisions with other planetesimals. Several of these changes can clearly be seen by comparison between the top and the bottom plots of each figure. For example, in Figure 6 the last visible step in the growth of the planetesimal coincides with the change from tadpole to horseshoe orbit.

\begin{figure}
\includegraphics[scale=.45]{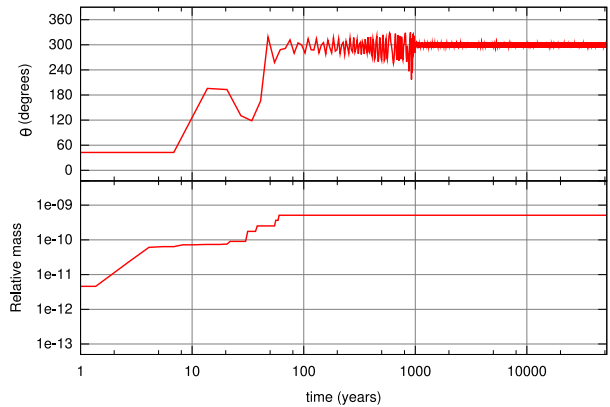}
\caption{Temporal evolution of the libration angle (top) and the relative mass. (bottom) of the remaining planetesimal from the simulation of a cloud of $10^4$ planetesimals with $m_{p_i}=10^{-13}$, initially distributed around $L_4$. 
After about 60 years the planetesimal reaches its final mass, $m_{p_f}=5.117\times10^{-10}$. and starts to librate around $L_5$.
}
\end{figure}

\begin{figure}
\includegraphics[scale=.45]{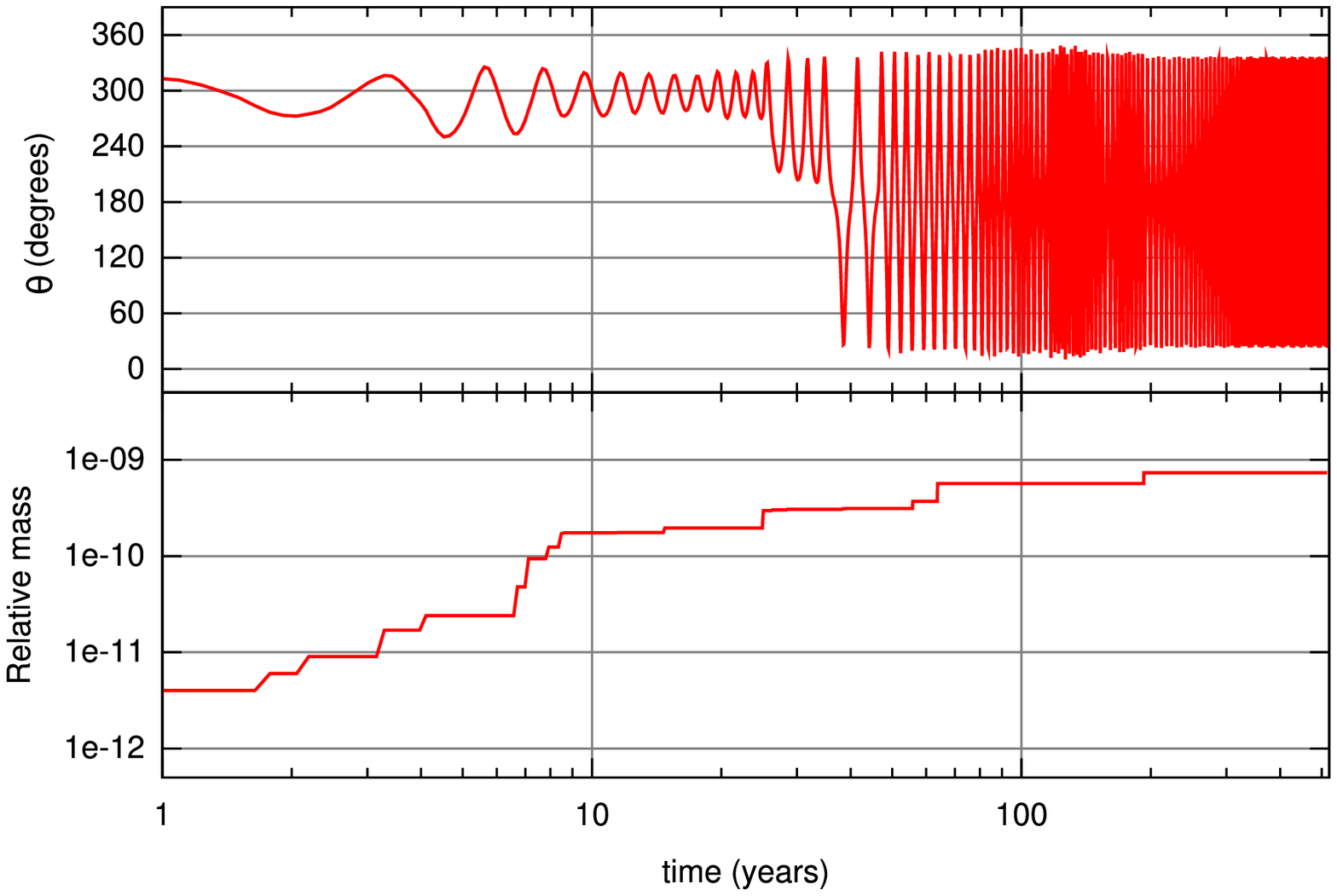}
\caption{Temporal evolution of the libration angle (top) and the relative mass. (bottom) of the remaining planetesimal from the simulation of a cloud of $10^3$ planetesimals with $m_{p_i}=10^{-12}$, initially distributed around $L_5$. 
By the end of the simulation the planetesimal reaches its final mass, $m_{p_f}=7.35\times10^{-10}$ and stays in a horseshoe orbit.
}
\end{figure}

\begin{figure}
\includegraphics[scale=.45]{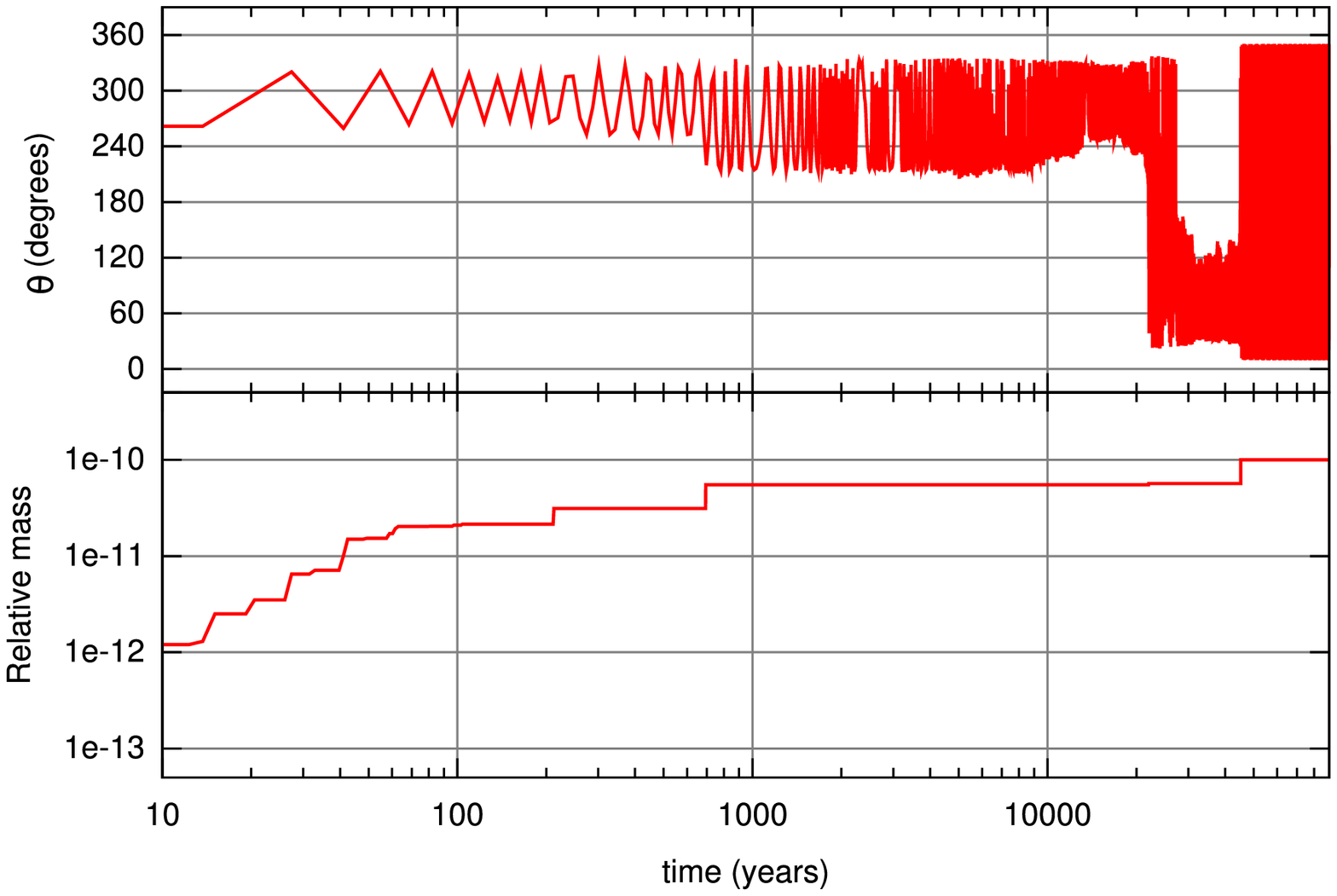}
\caption{Temporal evolution of the libration angle (top) and the relative mass. (bottom) of the remaining planetesimal from the simulation of a cloud of $10^3$ planetesimals with $m_{p_i}=10^{-13}$, initially distributed around $L_4$.
By the end of the simulation the planetesimal reaches its final mass, $m_{p_f}=10^{-10}$ and stays in a horseshoe orbit.
}
\end{figure}

\begin{figure}
\includegraphics[scale=.45]{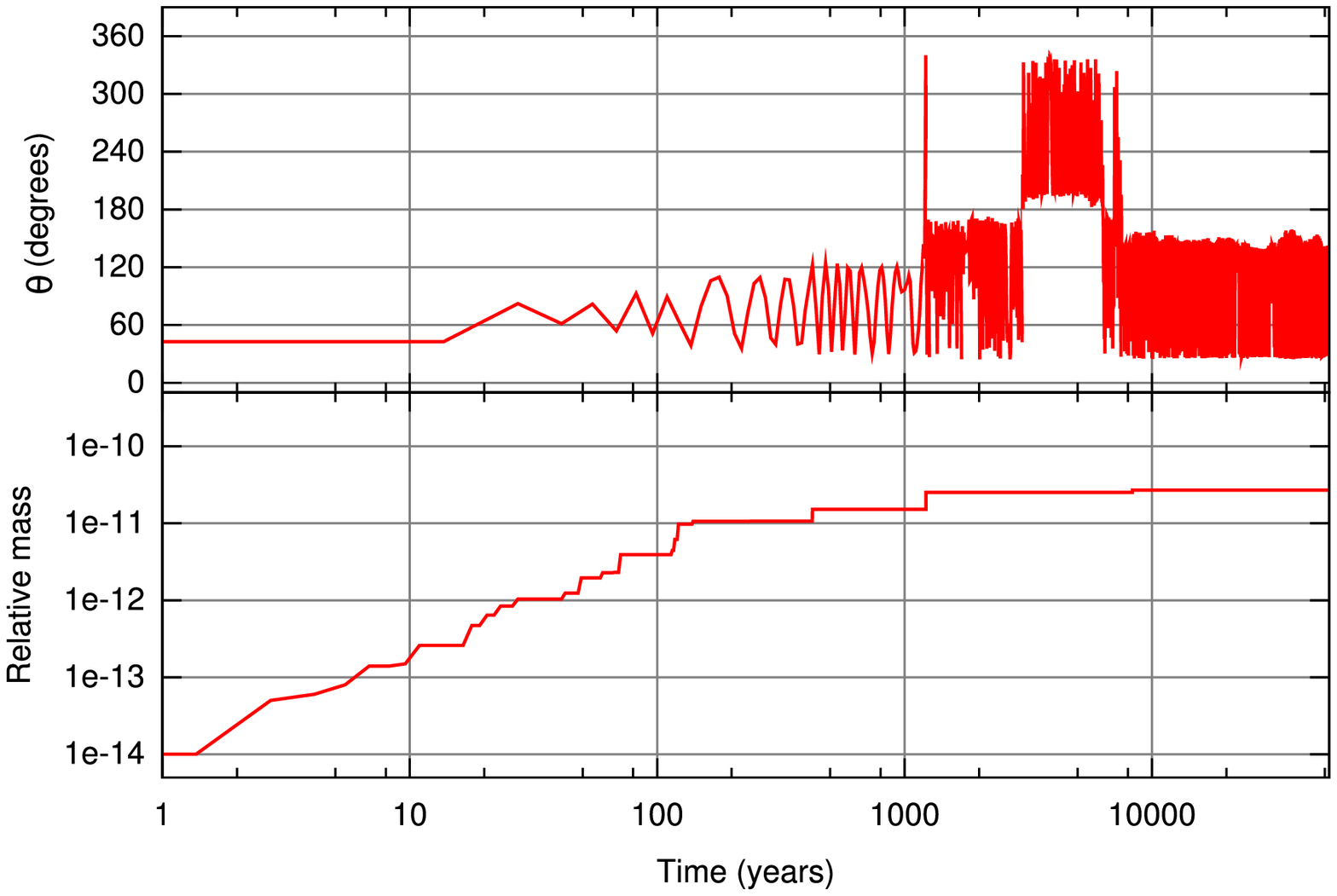}
\caption{Temporal evolution of the libration angle (top) and the relative mass. (bottom) of the remaining planetesimal from the simulation of a cloud of $5\times10^3$ planetesimals with $m_{p_i}=10^{-14}$, initially distributed around $L_4$. After about 51 thousand years the planetesimal reaches its final mass, $m_{p_f}=2.703\times10^{-11}$ and starts to librate in a tadpole orbit around $L_4$.
} 
\end{figure}

A summary of the whole set of simulations are presented in Tables 3 and 4. We performed 24 simulations neglecting the oblateness of Saturn (Table 3) and other 18 simulations taking it into account (Table 4). In the last two columns the kind of coorbital trajectory and the final mass of the survivors planetesimals are shown. As we can see, in several simulations there were more then one planetesimal left.
In some cases we followed the simulations for longer times. For example, simulation 13a (Table 3) was integrated for more then $10^5$ years and no extra collision occurred along the second half of such integration time. In the case of simulation 14a (Table 3), the integration was extended to one million years and there were still two planetesimals left. One in horseshoe orbit and other in tadpole around $L_5$.

From Table 3 we have that 2/3 of the survivors are left in horseshoe trajectories, while the remaining 1/3 are equally divided in tadpole trajectories around $L_4$ and $L_5$. We also note that, in each simulation, the total mass of the survivors is very close to the corresponding total mass of initial cloud of planetesimals.
The difference between these two values is the mass of planetesimals that collided with the proto-satellite.

Comparing the simulations presented in Table 3 with those presented in Table 4, we can note that the oblateness of Saturn did not introduce any significant change in the results as a whole.

\begin{table*}

 \centering
 \begin{minipage}{173mm}
  \caption{Set of simulations without Saturn's oblateness}

\begin{tabular}{@{}lcccccc@{}}
  \hline
  \\
Simulation & Number of         & Location of       & Integration        & $m_{p_i}$ & Trajectories of & final masses            \\
	  & plantesimals   & initial distribution & time (years)   &  & the survivors\footnote[2]{Horseshoe (H), Tadpole around $L_4$ ($T_{L_4}$) and Tadpole around $L_5$ ($T_{L_5}$)}  &  $m_{p_f}$                             \\
  \hline\hline
1         & 10000          & $L_4$        &  10005    & $10^{-14}$   &   H      &    $10^{-10}$                \\
\hline
2         & 10000          & $L_5$  &  20238         & $10^{-14}$    &   H    &   $10^{-10}$                \\
\hline
3         & 10000          & $L_4$   & 1068          & $10^{-13}$     &   $T_{L_5}$    &   $5.117\times10^{-10}$           \\
\hline
4         & 10000          & $L_5$   & 99         &     $10^{-13}$    &   H    &    $9.999\times10^{-10}$         \\
\hline
5         & 10000          & $L_4$   & 51         & $10^{-12}$        &    H   &    $9.969\times10^{-9}$        \\
\hline
6         & 10000          &  $L_5$    &50         &   $10^{-12}$     &    H  &    $9.974\times10^{-9}$        \\
\hline
7         & 5000  & $L_4$ &  51000    & $10^{-14}$     & H, $T_{L_4}$    &  $2.297\times10^{-11}$, $2.703\times10^{-11}$  \\
\hline
8         & 5000      & $L_5$ &  51000 & $10^{-14}$ & H, $T_{L_5}$ &$2.53\times10^{-11}$, $2.47\times10^{-11}$     \\
\hline
9         & 5000          & $L_4$   & 3325    & $10^{-13}$     & H   & $4.961\times10^{-10}$           \\
\hline
10         & 5000          & $L_5$   & 989    &  $10^{-13}$     &  H   & $4.943\times10^{-10}$           \\
\hline
11         & 5000          & $L_4$    & 52   & $10^{-12}$       &   H   & $3.602\times10^{-9}$           \\
\hline
12        & 5000          &  $L_5$   & 61   &   $10^{-12}$      &    $T_{L_5}$   &  $3.369\times10^{-9}$        \\
\hline
13a         & 1000          & $L_4$  &  51000    & $10^{-14}$          & H, H, $T_{L_4}$     & $3.45\times10^{-12}$, $1.94\times10^{-12}$, $4.61\times10^{-12}$         \\
\hline
13b        & 1000          & $L_4$   & 51000     &   $10^{-14}$        &   $T_{L_4}$, H, H   & $2.54\times10^{-12}$, $2.21\times10^{-12}$, $5.25\times10^{-12}$          \\
\hline
14a         & 1000          & $L_5$   &  51000   & $10^{-14}$     & H, $T_{L_5}$, H, H  & $3.35\times10^{-12}$, $9.6\times10^{-13}$, $4.92\times10^{-12}$, $7.7\times10^{-13}$           \\
\hline
14b         & 1000          & $L_5$   & 51000     & $10^{-14}$         & H, H, H, $T_{L_5}$  & $1.51\times10^{-12}$, $5.03\times10^{-12}$, $2.14\times10^{-12}$, $1.33\times10^{-12}$        \\
\hline
15a         & 1000          & $L_4$   & 5100   & $10^{-13}$    & H, $T_{L_4}$ &   $5.38\times10^{-11}$, $4.62\times10^{-11}$     \\
\hline
15b         & 1000          & $L_4$   & 5100    &  $10^{-13}$         & $T_{L_4}$, H    &  $4.24\times10^{-11}$, $5.76\times10^{-11}$       \\
\hline
16a         & 1000          & $L_5$   & 5100 &  $10^{-13}$     & $T_{L_4}$, H, $T_{L_5}$    & $1.7\times10^{-12}$, $4.32\times10^{-11}$, $5.51\times10^{-11}$        \\
\hline
16b         & 1000          & $L_5$   & 5100    & $10^{-13}$      & H, $T_{L_5}$    &   $4.28\times10^{-11}$, $5.72\times10^{-11}$       \\
\hline
17a         & 1000          &  $L_4$  & 510    & $10^{-12}$        & $T_{L_4}$, H       & $6.99\times10^{-10}$, $2.52\times10^{-10}$         \\
\hline
17b         & 1000          & $L_4$   & 229    & $10^{-12}$              &  H          &   $9.32\times10^{-10}$      \\
\hline
18a         & 1000          & $L_5$  & 192    &  $10^{-12}$         &  H    &  $7.35\times10^{-10}$          \\
\hline
18b         & 1000          & $L_5$   &  277    & $10^{-12}$        &  H    & $10^{-9}$       \\
\hline
\end{tabular}
\end{minipage}

\end{table*}

\begin{table*}

 \centering
 \begin{minipage}{173mm}
  \caption{Set of simulations {\bf with Saturn's oblateness}}

\begin{tabular}{@{}lcccccc@{}}
  \hline
  \\
Simulation & Number of         & Location of       & Integration        & $m_{p_i}$ & Trajectories of & final masses            \\
	  & plantesimals   & initial distribution & time (years)   &  & the survivors\footnote[2]{Horseshoe (H), Tadpole around $L_4$ ($T_{L_4}$) and Tadpole around $L_5$ ($T_{L_5}$)}  &  $m_{p_f}$                             \\
  \hline\hline
1         & 10000          & $L_4$        &  51000    & $10^{-14}$   &   H, H     &   $1,612\times10^{-11}$, $8.388\times10^{-11} $                \\
\hline
2         & 10000          & $L_5$         &  14672    & $10^{-14}$    &   H    &   $10^{-10}$                \\
\hline
3         & 10000          & $L_4$   & 654        & $10^{-13}$     &   H    &   $8.666\times10^{-10}$           \\
\hline
4         & 10000          & $L_5$   & 218         &     $10^{-13}$    &   H    &    $9.822\times10^{-10}$         \\
\hline
5         & 10000          & $L_4$   & 55         & $10^{-12}$        &    H   &    $9.382\times10^{-9}$        \\
\hline
6         & 10000          &  $L_5$    &59         &   $10^{-12}$     &    H  &    $9.923\times10^{-9}$        \\
\hline
7         & 5000  & $L_4$ &  51000    & $10^{-14}$     & H, $T_{L_4}$    &  $2.646\times10^{-11}$, $2.354\times10^{-11}$  \\
\hline
8         & 5000      & $L_5$ &  51000 & $10^{-14}$ & H, $T_{L_4}$ &$2.119\times10^{-11}$, $2.881\times10^{-11}$     \\
\hline
9         & 5000          & $L_4$   & 2308    & $10^{-13}$     & H   & $4.926\times10^{-10}$           \\
\hline
10         & 5000          & $L_5$   & 1663    &  $10^{-13}$     &  H   & $5\times10^{-10}$           \\
\hline
11         & 5000          & $L_4$    & 95   & $10^{-12}$       &   H   & $4.445\times10^{-9}$           \\
\hline
12        & 5000          &  $L_5$   & 49   &   $10^{-12}$      &    H   &  $4.333\times10^{-9}$        \\
\hline
13        & 1000          & $L_4$  &  51000    & $10^{-14}$          & H, H,H, $T_{L_4}$     & $3.18\times10^{-12}$, $3.45\times10^{-12}$, $1.79\times10^{-12}$,$1.58\times10^{-12}$         \\
\hline
14         & 1000          & $L_5$   &  51000   & $10^{-14}$     & H, H, $T_{L_5}$ & $4.39\times10^{-12}$, $2.12\times10^{-11}$, $3.49\times10^{-11}$          \\
\hline
15         & 1000          & $L_4$   & 29768   & $10^{-13}$    & H &   $6.78\times10^{-11}$     \\
\hline
16         & 1000          & $L_5$   & 51000 &  $10^{-13}$     & $T_{L_5}$, H    & $4.37\times10^{-11}$, $5.63\times10^{-11}$        \\
\hline
17         & 1000          & $L_4$   & 218    & $10^{-12}$              &  H          &   $9.86\times10^{-10}$      \\
\hline
18         & 1000          & $L_5$  & 173    &  $10^{-12}$         &  H    &  $9.99\times10^{-10}$          \\
\hline
\end{tabular}
\end{minipage}

\end{table*}

Beaugé et al. (2007) found that the final mass of the formed trojans the kind of terrestrial planets did not depend significantly neither on the initial mass, nor on the number of the initial population of planetesimals. In our simulations we do realize that the number of planetesimal and its initial mass play an important role in the final outcome.

In Table 3, we can see that the initial populations with small mass ($10^{-13}$,$10^{-14}$) and small number of planetesimal (1000) have a larger tendency to yield more than one final body. This effect is caused  by the weak gravitational attraction when the planetesimals are smaller and more distant from each other, consequently few collisions occur. However, when we have a lot of planetesimals, even with small mass, they are closer, then the gravitational attraction is larger and collisions become more frequent.

Although in most cases the final total mass is almost equal to the initial value, in one case there is a significant mass loss (Table 3, simulation 3). This occurred due to a collision of a larger planetesimal with the secondary body. However, for a similar initial planetesimal population  distributed around $L_5$ there is no mass loss, this loss mass might be one statistical fluke, but this has to be checked with further studies.
  
The temporal evolution of the total number of the remaining planetesimals, $N$, in the system gives an idea of the different speeds at each stage of the system evolution.
In Figure 8  we present a representative sample of the 
temporal evolution of the total number of the remaining planetesimals of our simulations.
In all the cases studied, the evolution of the system can be well represented by three stages. In the first stage there is a "cold" cloud, the planetesimals are in near circular orbits and do not collide frequently. Then, along the time the gravitational interaction among the planetesimals "heat" the cloud, the trajectories get some eccentricity and the rate of collisions increase, producing larger planetesimals. Larger planetesimals excite the cloud even more and also increase their sphere of influence, resulting in larger cross sections and faster growing. When most of the planetesimals collided, there are a few left, then the collision rate decreases significantly and the evolution of the system slows down.

A comparison of the curves presented in Figure 8 shows that the rate of decay of the total number of planetesimals is strongly dependent on the total mass of the initial cloud of planetesimals.

\begin{figure}
\includegraphics[scale=.6]{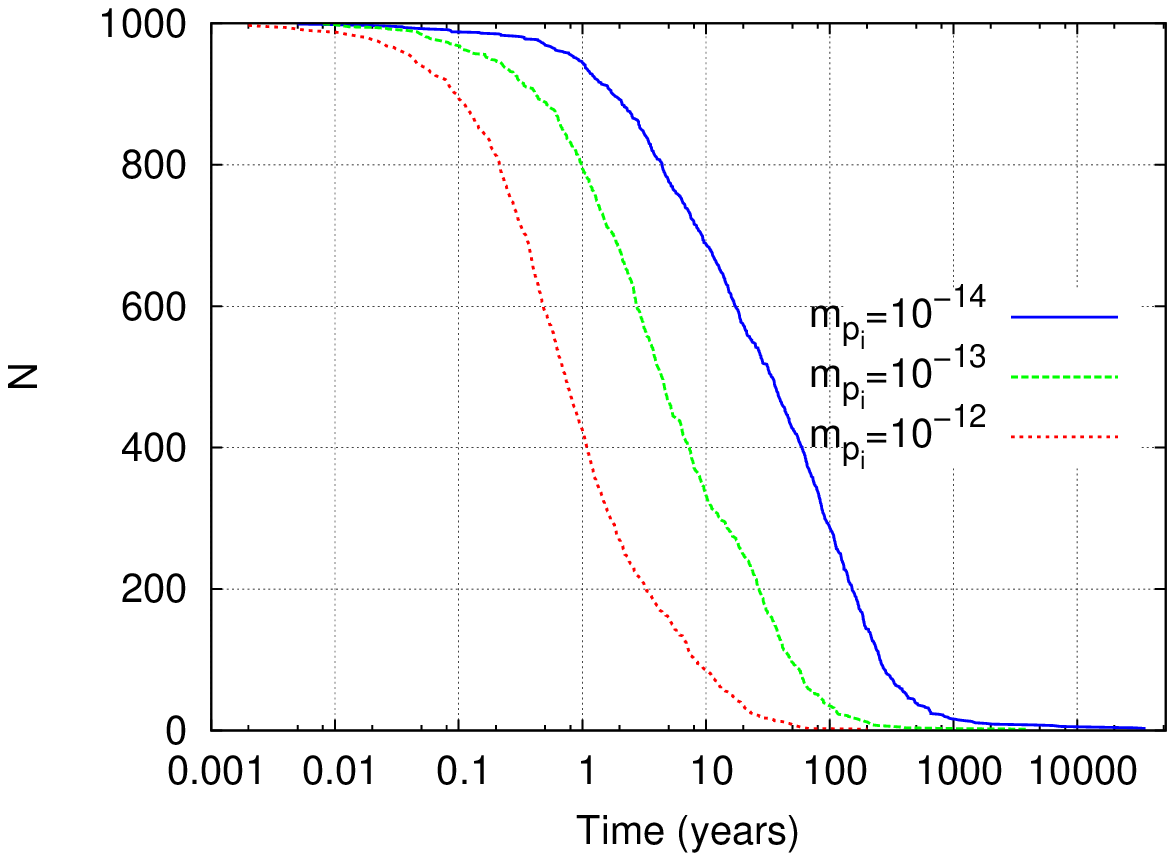}
\includegraphics[scale=.6]{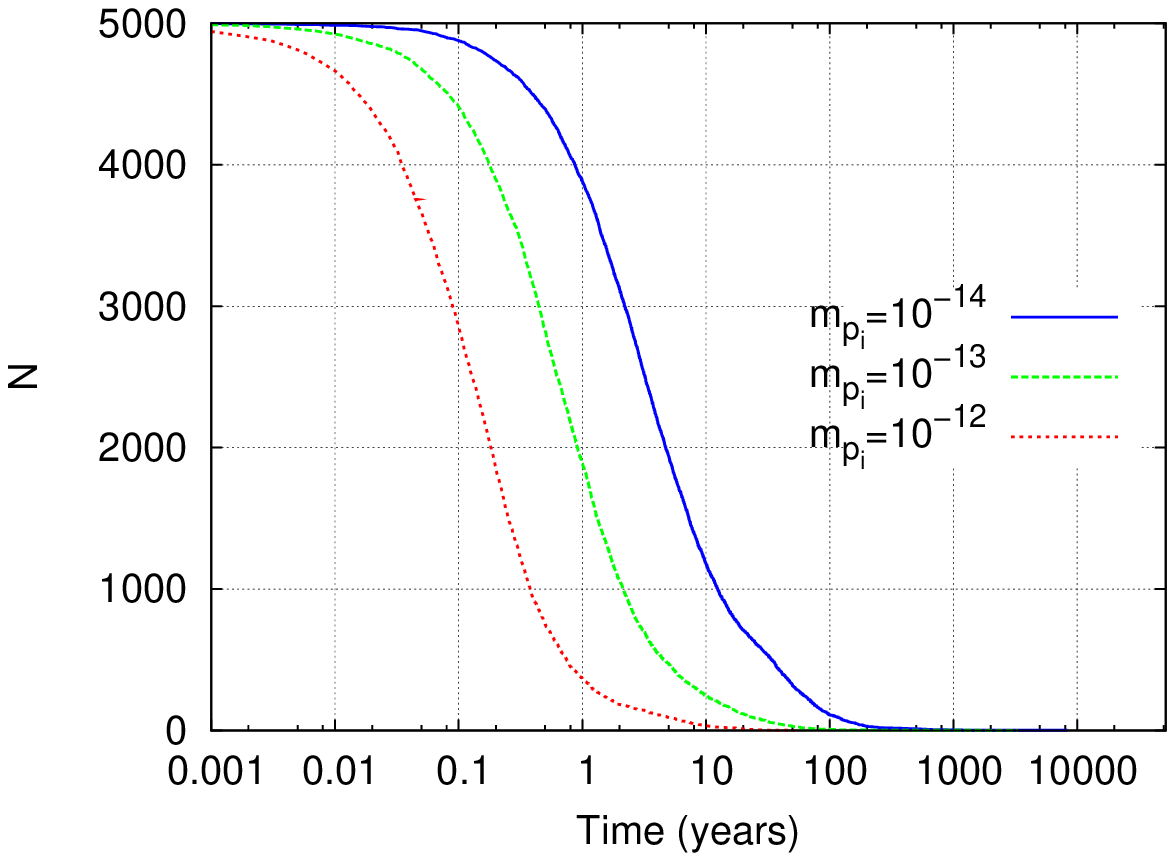}
\includegraphics[scale=.6]{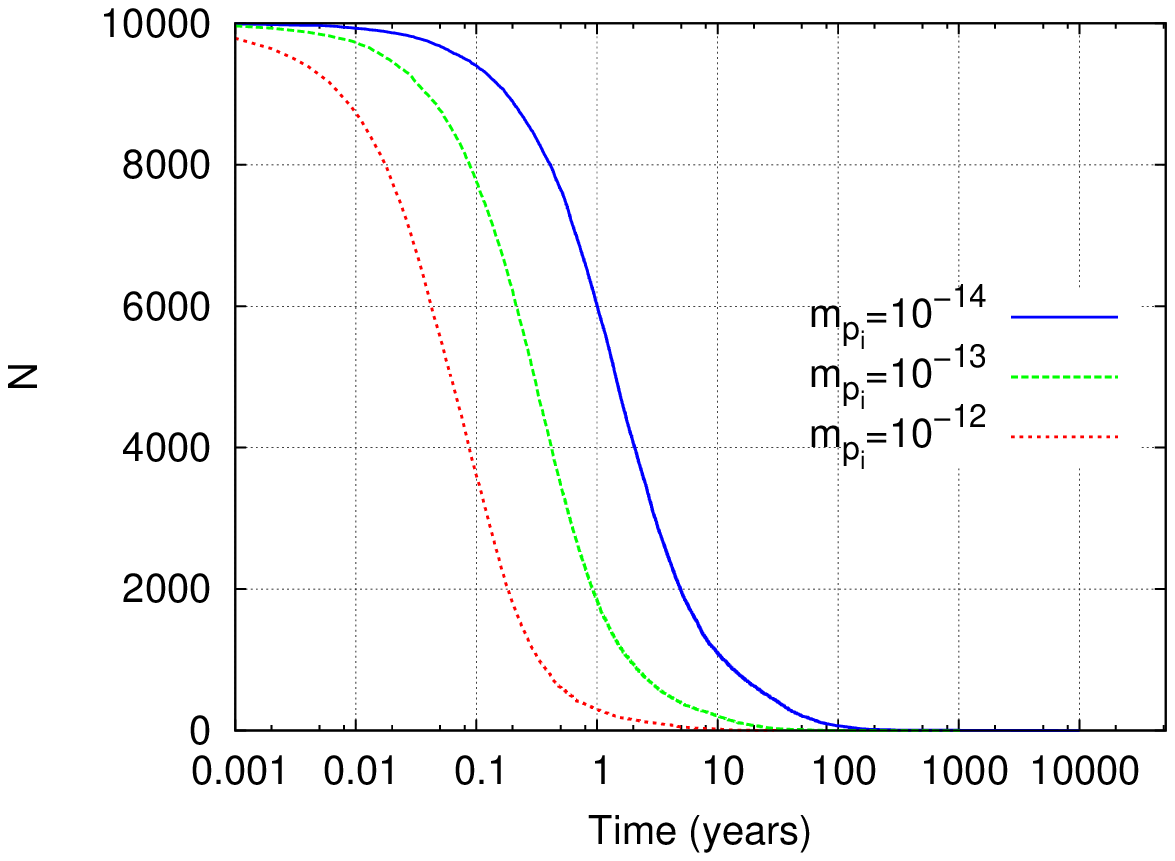}
\vspace{1.0cm}
\caption{Temporal evolution of the total number of remaining planetesimals.
It is shown the data from simulations of clouds of $10^3$ (top), $5\times 10^3$ (middle) and $10^4$ (bottom)  planetesimals around $L_4$
with initial masses $m_{p_i}=10^{-14}$, $m_{p_i}=10^{-13}$ and $m_{p_i}=10^{-12}$ }
\end{figure}

In order to infer possible effects due to mean motion resonances with nearby satellites on the mechanism of coorbitals formation, we performed simulations considering  Mimas and Thetys in 4:2 resonance of the type-inclination. We considered a system composed of an oblate central body (Saturn), two satellites (Mimas-Thetys) which are in mean-motion resonance 4:2 and a cloud of planetesimals. The simulations were made for six cloud of 1000 planetesimal considering in each case $m_{p_i}$ equal to $10^{-12}$, $10^{-13}$ or $10^{-14}$ around $L_4$ and $L_5$. It is important to have in mind that the previous simulations were in a planar system, and when we considered Mimas and Thetys in a 4:2 inclination-type resonance, the problem becomes three-dimensional. In general, these new simulations did not show any significant difference in comparison with the simulations presented in table 4. In Figure 9 we present a comparison of the temporal decaying number of the planetesimals for the simulations of a cloud of $10^3$ planetesimals initially located around $L_4$ and with $m_{p_i}=10^{-12}$. It shows that only in the final stages (after reducing the number of planetesimals to about 30\%) the results present some notable difference. The results under resonant perturbation evolves more slowly than the system without such perturbation. Probably, this is only due to the fact that the system under resonant perturbation is in a three-dimensional  space, which reduces the chances of encounters between the planetesimals in comparison with the planar system case.

\begin{figure}
\includegraphics[scale=.6]{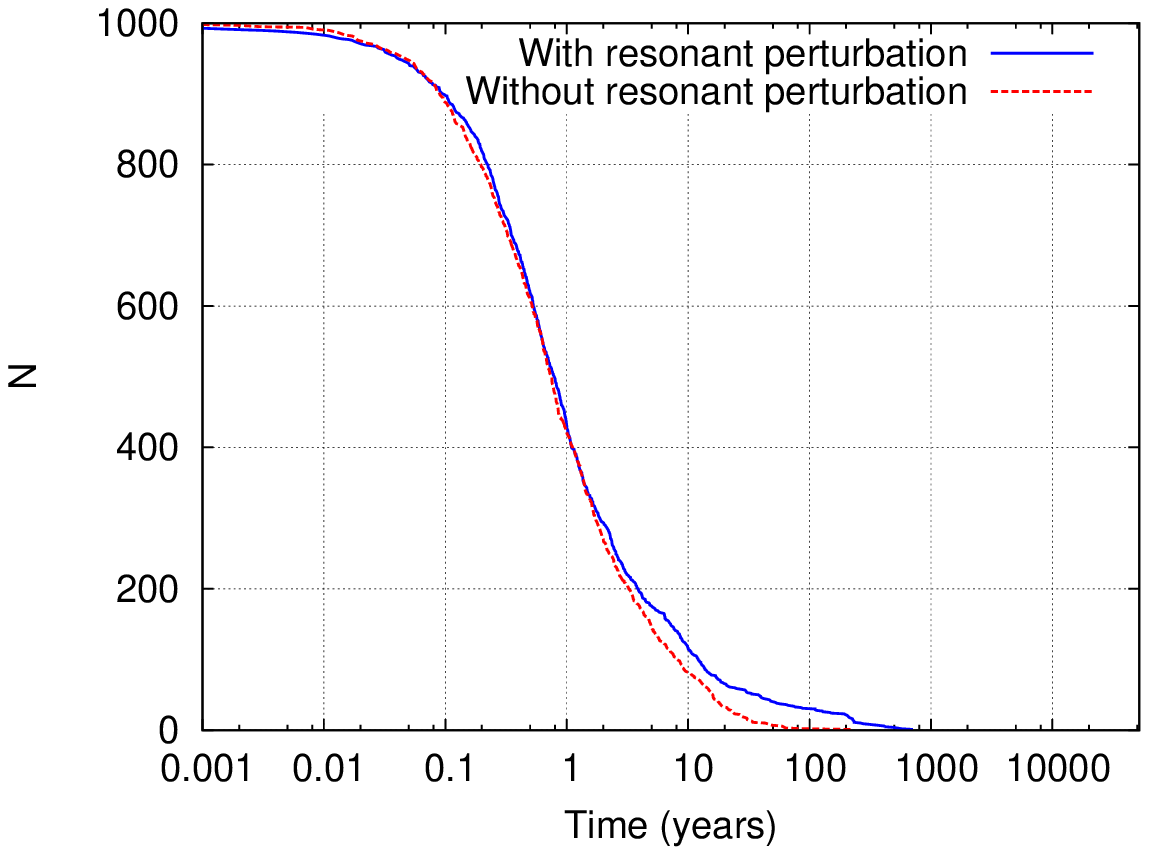}
\vspace{0.5cm}
\caption{Temporal evolution of the total number of remaining planetesimals. It is shown the data from simulations of clouds of $10^3$ with initial masses $m_{p_i}=10^{-12}$ initially distributed around $L_4$. The red and dashed line correspond to a simulation only with Thetys and the blue and solid line correspond to a simulation with Thetys and Mimas in mean motion resonance 4:2.}
\end{figure}

\section{Final Comments}
In the present work we present results of numerical simulations in order to study the viability of the congenital formation approach as a mechanism for the origin of the coorbital satellites of Thetys and Dione. We considered that the collisions are always inelastic and make the growth of the planetesimals. The results show that almost all the initial mass of the clouds are converted in one or a few planetesimals in coorbital motion with the proto-satellite. The final masses are of the same order of the small coorbital satellites Helene, Polydeuces, Telesto and Calypso. Therefore, we found that this is a promising mechanism for this purpose.

The distribution of librational amplitudes of the final bodies that remain in tadpole orbits varies from  4 up to 75 degrees.The real coorbitals have a libration amplitude of which varies from 1.3 up to 25.9 degrees. However, some mechanisms could reduce the amplitude of libration, for example, the gas drag (Chanut et al., 2008), mass accretion or radial migration of the secondary body (Fleming and Hamilton, 2000). In this paper we do not consider these mechanisms, but might be included in future works.

The effects of mean motion resonances of nearby satellites shows to be of no significative relevance for the present model of congenital formation of coorbitals.

The results presented here might look like obvious. However, in Beaugé et al (2007) it was found that a growing planetary embryo generates a chaotic region around the Lagrangian point, inhibiting additional accretion and setting a limit in the final mass of the coorbital body. They studied the possibility of congenital formation of Earth like planets in coorbital orbits of exoplanetary systems. In their work they adopted systems like the Jupiter-Sun system, with relative mass of $10^{-3}$, and did not manage to form coorbital planets with mass larger than $0.6M_\oplus$  This phenomena does not occur in the present work because the results of the simulation are strongly dependent on the relative mass of the secondary body (Izidoro et al., 2010).

It is important to comment that in the present congenital model the final coorbitals formed are of similar masses. However, Dione has one coorbital, Helene whose mass is about two orders of magnitude higher than the other coorbital, Polydeuces. Therefore, the model as considered in the present work is not able to explain by itself alone such configuration. Consequently, there is still plenty of work to be done.
\section*{Acknowledgments}
The comments and questions of an anonymous referee helped to improve this paper.
This work was supported by the Brazilian agencies CAPES and CNPq, and by the Foundation of the State of S\~ ao Paulo FAPESP.

\bsp

\label{lastpage}


\begin{thebibliography}{99}
\bibitem[\protect\citeauthoryear{Armitage}{2007}]{b1} Armitage P. J., 2007, arXiv:astro-ph/0701485
\bibitem[\protect\citeauthoryear{Beaugé et al}{2007}]{b2} Beaugé C., Sándor Zs., Érdi B., S\"{u}li Á., 2007, A\&A, 463, 359
\bibitem[\protect\citeauthoryear{Chambers}{1999}]{b3} Chambers J.,1999, MNRAS, 304, 793
\bibitem[\protect\citeauthoryear{Chanut et al.}{2008}]{b4} Chanut T., Winter O. C., Tsuchida M., 2008, A\&A, 481, 519
\bibitem[\protect\citeauthoryear{Christou et al.}{2007}]{b5} Christou A. A., Namouni F., Morais M. H. M., 2007, Icarus, 192, 106
\bibitem[\protect\citeauthoryear{Connors}{2005}]{b6} Connors, M., 2005, Planetary and Space Science, 53, 617
\bibitem[\protect\citeauthoryear{Dermott \& Murray}{1981}]{b7} Dermott S.F., Murray C.M., 1981a, Icarus, 48, 1
\bibitem[\protect\citeauthoryear{Dermott \& Murray}{1981}]{b8} Dermott S.F., Murray C.M., 1981b, Icarus, 48, 12
\bibitem[\protect\citeauthoryear{Fleming and Hamilton}{2000}]{b9} Fleming J. H., Hamilton D. P., 2000, Icarus, 148, 479
\bibitem[\protect\citeauthoryear{Giorgini et al.}{1996}]{b10} Giorgini, J.D., Yeomans, D. K., Chamberlin, A. B., Chodas, P. W., Jacobson, R. A., Keesey, M., Lieske,J. H., Ostro, S. J., Standish, E. M., Wimberly, R. N., 1996 ,BAAS, (28)
\bibitem[\protect\citeauthoryear{Greenberg}{1989}]{b11} Greenberg R., in Origin And Evolution of Planetary and Satellite Atmosphere (eds. Atreya, S. K., Pollack J. B., Matthews M. S.), Univ. Arizona Press, Tucson, 137
\bibitem[\protect\citeauthoryear{Hayashi et al.}{1985}]{b12} Hayashi C. K., Nakazawa K., Nakagawa Y., 1985, in Protostar and Planets II (eds. Black D. C., Matthews M. S.),
Univ. Arizona Press, Tucson, 1100
\bibitem[\protect\citeauthoryear{Innanen}{1991}]{b13}  Innanem K. A., 1991, J. R. Astron. Soc. Can., 85, 151
\bibitem[\protect\citeauthoryear{Izidoro}{2010}]{b31} Izidoro et al., in preparation, 2010
\bibitem[\protect\citeauthoryear{Jacobson}{2004}]{b31} Jacobson R. A.,Bull. Am. Astron. Soc, 36, 1097(2004)
\bibitem[\protect\citeauthoryear{Lecacheux et al.}{1980}]{b14} Lecauchex J., Laques P., Vapillon L., Auge A., Despiau R., 1980, Icarus, 43, 111 
\bibitem[\protect\citeauthoryear{Mourão et al.}{2006}]{b15} Mourão D. M., Winter O. C., Tokoyama T., Cordeiro R. R., 2006, MNRAS, 372, 1614 
\bibitem[\protect\citeauthoryear{Murray et al.}{2005}]{b16} Murray C. D., Cooper N. J., Evans M. W., Beurle K., 2005, Icarus, 179, 222
\bibitem[\protect\citeauthoryear{Porco et al.}{2005}]{b17} Porco C. C., Thomas P. C., Weiss J. W., Richardson D. C., 2005, Science, 307, 1226
\bibitem[\protect\citeauthoryear{Reitsema et al.}{1980}]{b18} Reitsema H. J., 1981, Icarus, 48, 140
\bibitem[\protect\citeauthoryear{Reitsema et al.}{1980}]{b19} Reitsema H. J., Smith B. A., Larson S. M., 1980, Icarus, 43, 116
\bibitem[\protect\citeauthoryear{Safronov}{1969}]{b20} Safronov V. S., 1969, Nauka Moscow, English Translation NASA TT F-677
\bibitem[\protect\citeauthoryear{Smith}{1981}]{b21} Smith B. A., Voyager Imaging Team, 1981, Science, 212, 504
\bibitem[\protect\citeauthoryear{Tabachnik \& Evans}{1999}]{b22} Tabachnik S., Evans N. W., 1999, Astrophys. J., 517, L63
\bibitem[\protect\citeauthoryear{Veillet}{1981}]{b23} Veillet C., 1981, A\&A, 102, L5
\bibitem[\protect\citeauthoryear{Wetherill}{1980}]{b24} Wetherill G. W., 1980, Ann. Rev. Astron. Astrophys., 18, 77
\bibitem[\protect\citeauthoryear{Wetherill}{1980}]{b25} Wetherill G. W., 1990, Ann. Rev. Earth Planet Sci., 18, 205
\bibitem[\protect\citeauthoryear{Wisdom}{1980}]{b26} Wisdom J., 1980, AJ, 85(8) 
//\bibitem[\protect\citeauthoryear{Yoder}{1979}]{b27} Yoder C. F., 1979, Icarus, 40, 341
\bibitem[\protect\citeauthoryear{Yoder}{1983}]{b28} Yoder C.F., Colombo G., Synnott S.P., Yoder K.A., 1983, Icarus 53, 431
\bibitem[\protect\citeauthoryear{Yoder \& Salo}{1988}]{b29} Yoder C.F., Salo H., 1988, A.A., 205,  309
\bibitem[\protect\citeauthoryear{Zhou et al.}{2008}]{b30} Zhou L.,  Dvorak R.,  Sun Y., 2008, arXiv:0811.2491v1 [astro-ph]












\end{thebibliography}
\end{document}